\documentclass{mn2e}
\usepackage{graphicx}
\usepackage{color}
\usepackage{psfrag}
\usepackage{subfigure}

\newcommand{\etal}{et~al.}

\newcommand{\cc}{$\mbox{cm}^{-3}$}
\newcommand{\kgcm}{$\mbox{kg\,m}^{-3}$}
\newcommand{\kmsMpc}{$\mbox{km\,s}^{-1}\,\mbox{Mpc}^{-1}$}

\newcommand{\WHz}{$\mbox{W\,Hz}^{-1}$}
\newcommand{\LGHz}{L_{\mbox{\tiny{1.4\,GHz}}}}
\newcommand{\Lradio}{L_{\mbox{\tiny{radio}}}}
\newcommand{\rCore}{r_{\mbox{\tiny{core}}}}
\newcommand{\rTrans}{r_{\mbox{\tiny{trans}}}}
\newcommand{\betaGalaxy}{\beta_{\mbox{\tiny{galaxy}}}}
\newcommand{\betaCluster}{\beta_{\mbox{\tiny{cluster}}}}
\newcommand{\Qjet}{Q_{\mbox{\tiny{jet}}}}
\newcommand{\sigmaQjet}{\sigma_{\mbox{\small{log\,Q{\tiny{jet}}}}}}
\newcommand{\nCore}{n_{\mbox{\tiny{core}}}}
\newcommand{\rhoCore}{\rho_{\mbox{\tiny{core}}}}
\newcommand{\RT}{R_{\mbox{\tiny{T}}}}
\newcommand{\Vmax}{V_{\mbox{\tiny{max}}}}
\newcommand{\Mbh}{M_{\mbox{\tiny{BH}}}}
\newcommand{\Mstar}{M_{\mbox{\small{$\star$}}}}
\newcommand{\tOn}{t_{\mbox{\tiny{on}}}}
\newcommand{\tOff}{t_{\mbox{\tiny{off}}}}
\newcommand{\Dn}{D_{\mbox{\small{n}}}}
\newcommand{\Mr}{M_{\mbox{\small{r}}}}

\newcommand{\MdotCool}{\dot{M}_{\mbox{\tiny{cool}}}}

\begin{document}

\title[The Duty Cycle of Local Radio Galaxies] 
{The Duty Cycle of Local Radio Galaxies}

\author[Shabala \etal\/]{S.S. Shabala$^\star$, S. Ash, P. Alexander, J.M. Riley\\
Astrophysics Group, Cavendish Laboratory, Madingley Road, Cambridge CB3 0HE, United Kingdom\\
Email: sshabala@mrao.cam.ac.uk}

\maketitle

\begin{abstract}

We use a volume- and flux-limited sample of local ($0.03 \leq z \leq 0.1$) radio galaxies with optical counterparts to address the question of how long a typical galaxy spends in radio-active and quiescent states. The length of the active phase has a strong dependence on the stellar mass of the host galaxy. Radio sources in the most massive hosts are also retriggered more frequently. The time spent in the active phase has the same dependence on stellar mass as does the gas cooling rate, suggesting the onset of the quiescent phase is due to fuel depletion. We find radio and emission line AGN activity to be independent, consistent with these corresponding to different accretion states. 

\end{abstract}


\section{Introduction}
\label{sec:introduction}

Recent theoretical and observational evidence suggests that the growth of galaxies and supermassive black holes at their centres are closely related phenomena (Magorrian et al. 1998; Gebhardt et al. 2000; H\"aring \& Rix 2004). As accretion onto these black holes is believed to power Active Galactic Nuclei (AGN) jets, such AGN activity is thus intricately linked to the process of galaxy formation and evolution. The notion of AGN feedback through radio sources, where the AGN jets heat up and expel the surrounding gas, has received particular attention in recent times for a number of reasons. Jet heating of the intracluster gas (ICM) decreases the rate of accretion onto the central black hole, until it is shut off completely. Once the gas has had sufficient time to cool, the accretion can restart. Although one can envisage a stable configuration in which the accretion rate and gas heating are balanced, outward transport of central gas by the jet (e.g. Basson \& Alexander 2003) implies that AGN heating must be sporadic in order to keep some gas available for fuelling \cite{KawataGibson05}. Even if all the fuel is carried away from the central regions, the AGN jet can restart once this gas is replenished, for example via an interaction with another galaxy \cite{BahcallEA97}. Such an interaction will eventually lead to an increase in the accretion rate and hence a re-launching of the radio jets. This inherent intermittency of the feedback process is confirmed by observations of rising radio bubbles \cite{ChurazovEA01} and multiple radio lobe pairs in the same source (Venturi \etal\/ 2004; Giovannini \etal\/ 1998, 1999). The intermittency also provides a natural way of keeping the black hole and spheroid growth in step. Furthermore, AGN heating has also been invoked to explain the lack of star formation in the most massive galaxies (the so-called cosmic downsizing; e.g. Croton \etal\/ 2006), and suppression of cooling flows in the cores of massive clusters \cite{FabianEA03,FormanEA05}. 

The questions of how long an average radio source spends in an active state, and the length of time between outbursts, are thus crucial to quantifying the importance of AGN feedback. A number of methods have been employed in attempts to quantify these. The oldest and best-known of these invokes spectral ageing and lobe expansion speed arguments \cite{AlexanderLeahy87}, yielding on-times of a few times $10^7$~years. Simulations suggest the jets will be disrupted after times of the order of $10^8$~years (Tucker \& David 1997; Omma \& Binney 2004); while energy injection rates required to quench cooling flows point to timescales of the order of few times $10^7$- few times $10^8$~years \cite{NulsenEA05,OwenEilek98,McNamaraEA05}. Observations of revived radio relic sources \cite{EnsslinGopalKrishna01} constrain the time between outbursts to around $10^9$~years. The local radio loud fraction of a few percent (e.g. Best \etal\/ 2005b) also suggests the radio sources spend an order of magnitude longer in their quiescent state than in the active state. 

Radio source evolution depends strongly on both the enviroment of the radio source, and jet characteristics \cite{KA97,KDA97,Alexander00,KaiserCotter02}. Recently, Best \etal\/ \shortcite{Paper2} have found a strong dependence of the local ($0.03 \leq z \leq 0.1$) radio loud AGN fraction on the stellar mass of the host galaxy. In other words, it appears that the radio jets get retriggered more frequently in massive hosts. Best \etal\/ (2005a,b) also investigated the relationship between radio and emission line AGN activity by splitting their radio-optical sample into AGNs and star-forming galaxies according to the diagnostics appropriate for each case, and found that these are different phenomena. This finding is confirmed by studies of radio and emission line AGN activity in the brightest group and cluster galaxies \cite{BestEA07}. In this work we investigate these intriguing results in more detail by employing a flux- and volume-limited sample of radio sources with host properties.

Much like Best \etal\/ \shortcite{Paper1}, we combine optical information available from the Sloan Digital Sky Survey (SDSS) Data Release 2 \cite{YorkEA00} with two radio surveys, NRAO VLA Sky Survey (NVSS; Condon \etal\/ 1998) and Faint Images of the Radio Sky at Twenty Cm (FIRST; Becker \etal\/ 1995) to construct a local ($0.03 \leq z \leq 0.1$) radio-optical sample. Unlike previous works, which are only flux-limited, however, our sample is both volume- and flux-limited at both the radio and optical (r-band) wavelengths. We also use a combination of the FIRST and NVSS 1.4~GHz fluxes, ensuring sensitivity to both compact and diffuse structures.

As the main focus of this work is on radio AGN activity, we split our sample into AGNs and star-forming galaxies according to the radio diagnostic, and construct the bivariate radio-optical luminosity function by further dividing the AGN subsample in stellar mass. Radio source models are then used to fit individual source sizes and luminosities (as derived from the FIRST and NVSS catalogues) and derive the typical length of the active phase as a function of stellar mass. Observed bivariate luminosity functions then set the time a radio source spends in a quiescent state.

The paper is structured as follows. In Section~\ref{sec:sample} we describe our volume- and flux-limited sample. Separation into AGNs and star-forming galaxies, as well as the resultant radio and bivariate luminosity functions, are discussed in Section~\ref{sec:RLF}. Section~\ref{sec:radioSourceModel} outlines our radio source model, and the results are presented in Section~\ref{sec:results}. We conclude with a discussion of our findings in the context of AGN fuelling mechanisms in Section~\ref{sec:discussion}.

Throughout the paper we assume the cosmological parameters $\Omega_M = 0.3$, $\Omega_\Lambda = 0.7$ and $H_0 = 70$~\kmsMpc.

\section{Sample}
\label{sec:sample}

The sample used in this work was obtained by cross-correlating the optical Sloan Digital Sky Survey (SDSS) Data Release 2 \cite{YorkEA00} with two Very Large Array (VLA) radio surveys at 1.4~GHz, the NRAO VLA Sky Survey (NVSS; Condon \etal\/ 1998), and Faint Images of the Radio Sky at Twenty Centimetres (FIRST; Becker \etal\/ 1995). The SDSS catalogue used here derives from the volume-limited ($0.005 \leq z \leq 0.1$) sample described by Nikolic \etal\/ \shortcite{NikolicEA04}. The sample covers a spectroscopic area of 2627 deg$^2$ on the sky and is optically complete, containing 44,630 galaxies with Petrosian r-band magnitude $\Mr<-20.45$. For a complete description of the sample and various derived optical parameters the reader is referred to Nikolic \etal\/ \shortcite{NikolicEA04}.

\subsection{Catalogue construction}
\label{sec:catalogueConstruction}

The two radio surveys used in constructing our radio-optical sample complement each other, in the sense that NVSS is more sensitive to diffuse sources, but provides less accurate position estimates (resolution of $\sim 45$~arcsec compared to $5$~arcsec for FIRST). The process of pairing the radio and optical catalogues by comparing celestial positions of individual objects was performed using the Catalogue Utilities for Reporting, Selecting and Arithmetic (CURSA) software package produced by the Starlink Project. Each of the radio surveys was independently paired with SDSS, and the resultant catalogues then paired together to eliminate any duplicates.

Crucial to the pairing process is the separation distance within which two identifications are assumed to correspond to the same object. Two different methods were employed to address this issue. In the first method, we followed Ivezi\'c \etal\/ \shortcite{IvezicEA02} and performed the pairing process at various separations, ranging from 1-40 arcsec for FIRST/SDSS, and 1-60 arcsec for NVSS/SDSS. The number of matches was plotted as a function of separation distance, and the point at which the number of matches stays approximately unchanged for any further increase in separation distance was taken as the best pairing distance estimate. This yielded a separation distance of 3~arcsec for FIRST/SDSS, and 12-15~arcsec for NVSS/SDSS. The second method involved generating a mock catalogue by offsetting all sources in the SDSS catalogue by 4 {\it arcmin} in right ascention. Both the mock and the actual SDSS catalogues were then paired with FIRST and NVSS for a range of separation distances. At the largest separation distance, we expect all real matches to be captured, plus an unknown number of false matches (due to the allowed separation distance being too large). An approximate number of real matches was obtained by subtracting the number of total matches for the real catalogue from the corresponding value for the mock catalogue. The summary of ``true'' and ``false'' matches thus obtained is given as a function of separation distance in Table~\ref{tab:catalogueMatching}. The optimum separation distance yields the highest possible completeness rate (given by the fraction of real matches out of all matches), and lowest contamination (fraction of false matches). We thus chose separation distances of 5~arcsec for the FIRST/SDSS pairing process (giving 2448 matches), and 15~arcsec for NVSS/SDSS (giving 2393 matches).

The two catalogues thus obtained were paired together with a separation distance of 0.1 arcsec, the level of SDSS coordinate accuracy. There were 1427 sources common to both catalogues; 1021 FIRST/SDSS objects did not have NVSS/SDSS counterparts, while 966 NVSS/SDSS sources were without corresponding identifications in the FIRST/SDSS catalogue.

\begin{table*}
\small
\begin{tabular}{|l||c|c||c|c|} \hline
{\bf FIRST/SDSS} & & & & \\ \hline
{Separation}		& {Real matches}     & {Completeness}	& {False matches} & {Contamination}	\\ \hline
1.5 arcsec		        & 2243		& 87.7\%		& 0		& 0\%		\\
3 arcsec		        & 2391		& 93.4\%		& 2		& 0\%		\\
5 arcsec		        & 2448 		& 95.7\%		& 13		& 0.5\%		\\
10 arcsec		        & 2553  	& 99.8\%		& 77		& 3.0\%		\\
20 arcsec		        & 2814		& 100\%			& 289		& 11.3\%	\\
30 arcsec		        & 3221		& 100\%			& 651		& 25.4\%	\\
40 arcsec		        & 3745		& 100\%			& 1186		& 46.3\%	\\ \hline
\hline \\
{\bf NVSS/SDSS} & & & & \\ \hline
{Separation}		& {Real matches} & {Completeness}	& {False matches} & {Contamination}	\\ \hline
5 arcsec		        & 1205 		& 43.5\%		& 15		& 1.2\%		\\
10 arcsec		        & 2001  	& 72.2\%		& 62		& 3.1\%		\\
12 arcsec		        & 2200		& 79.4\%		& 81		& 3.7\%		\\
15 arcsec		        & 2393		& 86.3\%		& 119		& 5.0\%		\\
20 arcsec		        & 2656	 	& 95.8\%		& 202		& 7.6\%		\\
40 arcsec		        & 3600		& 100\%			& 880		& 24.4\%	\\
60 arcsec		        & 4799		& 100\%			& 2027		& 42.2\%	\\ \hline
\end{tabular}
\caption{FIRST/SDSS and NVSS/SDSS catalogue matching summary. For FIRST/SDSS we adopt 3~arcsec as the separation distance with the best compromise between maximum completeness and minimum contamination. For NVSS/SDSS the corresponding value is 15~arcsec.}
\label{tab:catalogueMatching}
\end{table*}

There are two reasons for this. FIRST covers less of the sky than NVSS, meaning some NVSS/SDSS sources will not have a corresponding FIRST/SDSS match. The rest of the mismatches come about due to the differences in sensitivity, positional accuracy and resolution of the two radio surveys. For example, FIRST has higher resolution, which means nearby sources confused in NVSS are often resolved in FIRST. Conversely, diffuse radio structures with no bright central core are not picked up by FIRST, but appear in NVSS. All the mismatches in areas of the sky common to FIRST and NVSS were therefore examined visually, and eventually all of these were included in the catalogue, yielding 3110 sources of NVSS/FIRST/SDSS detections in the volume $0.005 \leq z \leq 0.1$.

\begin{table*}
\small
\begin{tabular}{|c|l||c|l|}
\hline
\multicolumn{2}{|c||}{Use FIRST Flux} & \multicolumn{2}{c|}{Use NVSS Flux} \\ \hline
F1	& source is compact in FIRST 		& N1	& source is compact in FIRST \\
	& (possibly extended in NVSS) 		&	& (possibly extended in NVSS) \\
	& and there are other compact 		&	& with no other obvious unrelated \\
	& sources in the NVSS envelope 		&	& sources in the NVSS envelope  \\
	& or a source with no obvious 		&	& \\
	& connection to the central 		&	& \\
	& FIRST/SDSS source of interest		&	& \\
F2 	& no catalogued source in NVSS 		& N2	& source is extended in NVSS \\
	& because either the NVSS field 	&	& or unresolved and will need \\
	& is empty or the flux detected by 	&	& integrating \\
	& NVSS is too low to be officially 	&	& \\
	& identified as a source		&	& \\
F3	& source is confused in NVSS 		& N3	& source is extended in NVSS \\
	& but potentially extended in 		&	& within the FIRST envelope \\
	& FIRST and will need integrating	&	& with no other unrelated sources \\
	&					&	& in either field and NVSS flux is  \\
	&					&	& complete \\
F4	& calculated integrated flux is 	& N4	& source not detected in FIRST \\
	& higher in FIRST and/or assumed 	&	& because emission was too \\
	& to be more accurate			&	& diffuse and/or extended \\ \hline
\end{tabular}
\caption{Flux criteria.}
\label{tab:fluxCriteria}
\end{table*}

Appropriate radio source fluxes were determined for each source using the criteria of Table~\ref{tab:fluxCriteria}. We note that, in contrast to Best \etal\/ \shortcite{Paper1}, we do not always use the NVSS flux for sources which are identified in both radio surveys. This ensures exclusion of contributions from unrelated FIRST sources that appear as a single confused source in the NVSS envelope. The catalogue was truncated to exclude all sources below the NVSS sensitivity limit of 3.4~mJy (corresponding to 99\% completeness in NVSS, the less sensitive of the two radio surveys; Condon \etal\/ 1998). A lower redshift limit of $z=0.03$ was imposed in order to ensure accurate positional matching of the catalogues. This yielded the final volume- and flux-limited sample of 1191 radio sources with optical identifications.

\subsection{Selection effects}
\label{sec:selectionEffects}

The use of two radio surveys with complementary features ensures the sensitivity of the sample to both compact and extended structure. FIRST will resolve confused NVSS sources into multiple compact components. NVSS, on the other hand, picks up diffuse structures that FIRST misses. There are, however, some types of sources that the approach outlined above is not sensitive to. One such example includes the very extended NVSS sources that would only be identified as cores by FIRST/SDSS. This problem is circumvented by visual inspection of all 1191 radio sources in the final catalogue.

More troublesome are sources with no bright central component in either of the two radio surveys. If the extended components (such as lobes) of these objects are further away from the centre of the optical galaxy than the NVSS/SDSS separation distance of 15~arsec, they are missed by the pairing process. Hence the presented catalogue does not include diffuse (greater than a few tens kpc in the adopted cosmology) radio relics. Best \etal\/ (2005a) constructed a similar SDSS/NVSS/FIRST catalogue (albeit using slightly different flux criteria and without the r-band magnitude cutoff), but adopted a more sophisticated pairing process. These authors used a search radius of 3~arcmin (corresponding to a few hundred kpc in the adopted cosmology) to identify multi-component NVSS sources. They found 6\% of their sample to consist of such objects. Since these include sources with radio cores that {\it are} picked up by the pairing process employed here, missing diffuse sources without compact cores is not expected to alter the conclusions of this work significantly. Their relatively small number is consistent with the notion that radio relics fade away quickly once energy supply to the hotspots ceases \cite{GiovanniniEA99}.

\section{The local radio luminosity function}
\label{sec:RLF}

\subsection{Observed luminosity function}
\label{sec:observedRLF}

\begin{table*}
\begin{centering}
\small
\begin{tabular}{c | rc | rc | rc} \hline
  \multicolumn{1}{c}{} & \multicolumn{2}{c}{All radio sources} & \multicolumn{2}{c}{radio loud AGNs} & \multicolumn{2}{c}{Star-forming galaxies} \\ \hline
  $\log \left( \frac{L_{\rm 1.4}}{\rm W \, Hz^{-1}} \right)$ & $N$ & $\log \Phi$ & $N$ & $\log \Phi$ & $N$ & $\log \Phi$ \\ \hline
  21.90 &  52 & $-3.91^{+0.08}_{-0.10}$ &   7 & $-4.98^{+0.18}_{-0.30}$ &  45 & $-3.96^{+0.09}_{-0.11}$ \\
  22.30 & 249 & $-3.99^{+0.03}_{-0.04}$ &  63 & $-4.60^{+0.06}_{-0.08}$ & 186 & $-4.12^{+0.04}_{-0.04}$ \\
  22.70 & 379 & $-4.50^{+0.03}_{-0.03}$ & 166 & $-4.91^{+0.04}_{-0.04}$ & 213 & $-4.73^{+0.04}_{-0.04}$ \\
  23.10 & 272 & $-4.99^{+0.03}_{-0.03}$ & 196 & $-5.13^{+0.03}_{-0.04}$ &  76 & $-5.54^{+0.05}_{-0.06}$ \\
  23.50 &  95 & $-5.46^{+0.05}_{-0.05}$ &  82 & $-5.52^{+0.05}_{-0.06}$ &  13 & $-6.32^{+0.12}_{-0.17}$ \\
  23.90 &  74 & $-5.57^{+0.05}_{-0.06}$ &  71 & $-5.59^{+0.06}_{-0.06}$ &   3 & $-6.96^{+0.22}_{-0.48}$ \\
  24.30 &  34 & $-5.92^{+0.08}_{-0.10}$ &  33 & $-5.93^{+0.08}_{-0.10}$ &   1 & $-7.45^{+0.28}_{-1.00}$ \\
  24.70 &  21 & $-6.13^{+0.10}_{-0.13}$ &  20 & $-6.15^{+0.10}_{-0.13}$ &   1 & $-7.46^{+0.28}_{-1.00}$ \\
  25.10 &  14 & $-6.32^{+0.12}_{-0.16}$ &  14 & $-6.32^{+0.12}_{-0.16}$ & $-$ & $-$ \\
  25.50 &   1 & $-7.47^{+0.33}_{-1.00}$ &   1 & $-7.47^{+0.28}_{-1.00}$ & $-$ & $-$ \\ \hline
\end{tabular}
\caption{Local radio luminosity function at 1.4 GHz for all sources, and split up into AGNs and star-forming galaxies using the 4000\AA\/ break classification. Number density $\Phi$ is in units of $L_{\rm 1.4}^{-1}$~Mpc$^{-3}$.}
\label{tab:RLF}
\end{centering}
\end{table*}

\begin{figure}
\centering
  \psfrag{xLabel}{$\LGHz\/$ / \WHz\/}
  \includegraphics[height=0.48\textwidth,angle=270]{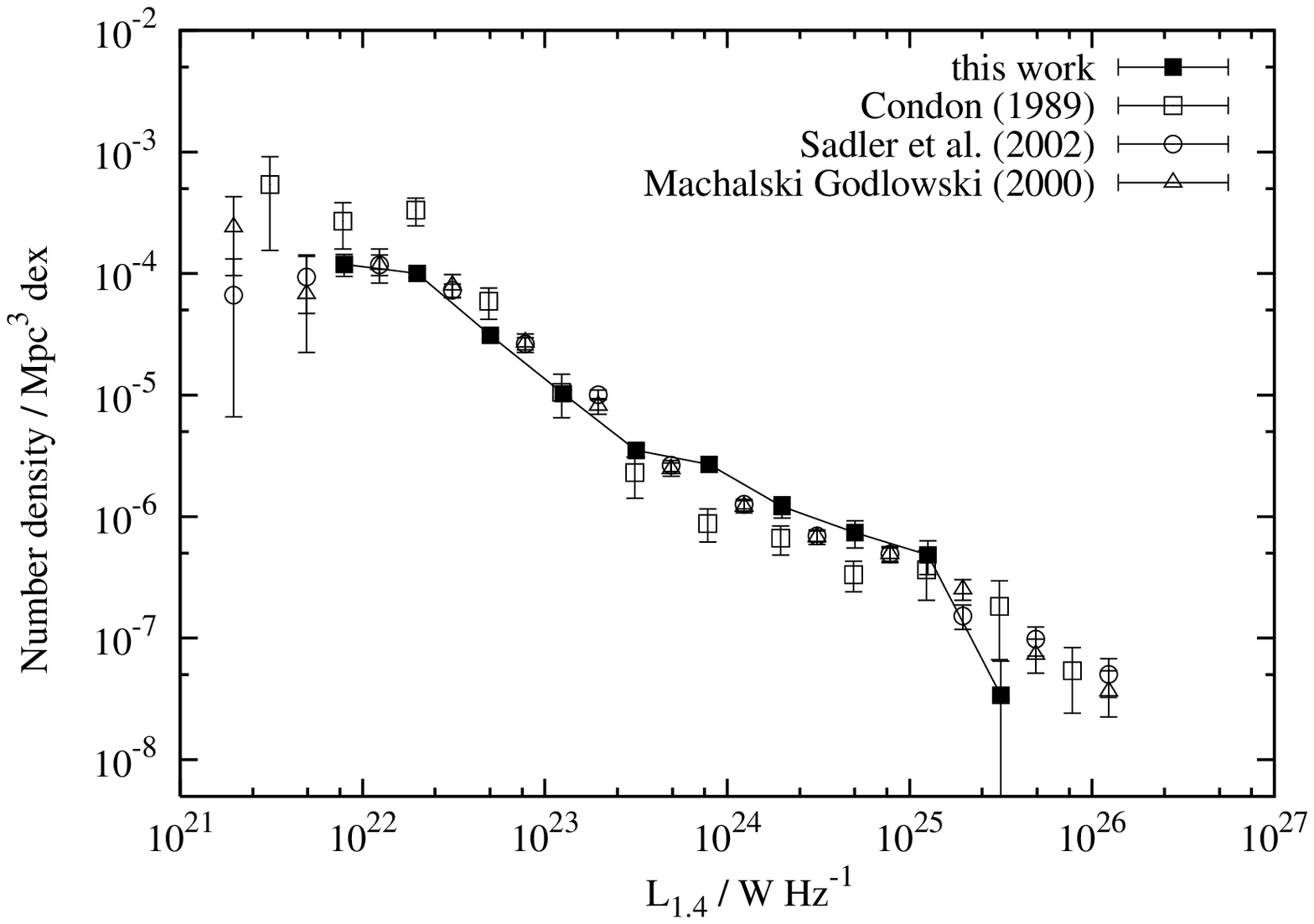}
\label{fig:RLFall}
\caption{Local radio luminosity function at 1.4 GHz for all sources. It is in good agreement with other studies (Sadler \etal\/ 2002 using 2dFGRS, Machalski \& Godlowski 2000 using the LCRS, and Condon 1989 using a VLA sample). We also find similar good agreement at high ($\LGHz\/ > \sim 10^{23}$~\WHz\/) luminosities with Best \etal\/'s \protect\shortcite{Paper1} RLFs for a flux-limited NVSS/SDSS sample. To compare these directly we averaged our luminosity functions over one dex in $\LGHz\/$, instead of the bin width of 0.3 dex. At $\LGHz\/ < \sim 10^{23}$~\WHz\/ we significantly underestimate the source counts when compared with Best \etal\, due to our sample being volume-limited.}
\end{figure}

We construct the radio luminosity function (RLF) for our radio flux- and volume-limited sample. In our cosmology, sources with $\LGHz < 8 \times 10^{22}$~\WHz\/ are too weak to be detected throughout the observed volume, and are only seen because of their low redshifts. For these sources we apply the usual $V/\Vmax\/$ correction \cite{Condon89}. The resultant RLF is given in Table~\ref{tab:RLF} and also plotted in Figure~\ref{fig:RLFall}, along with relevant previous works.

\begin{figure*}
\centering
  \psfrag{xLabel}{$\LGHz\/$ / \WHz\/}
  \subfigure[AGNs]{\includegraphics[height=0.4\textwidth,angle=270]{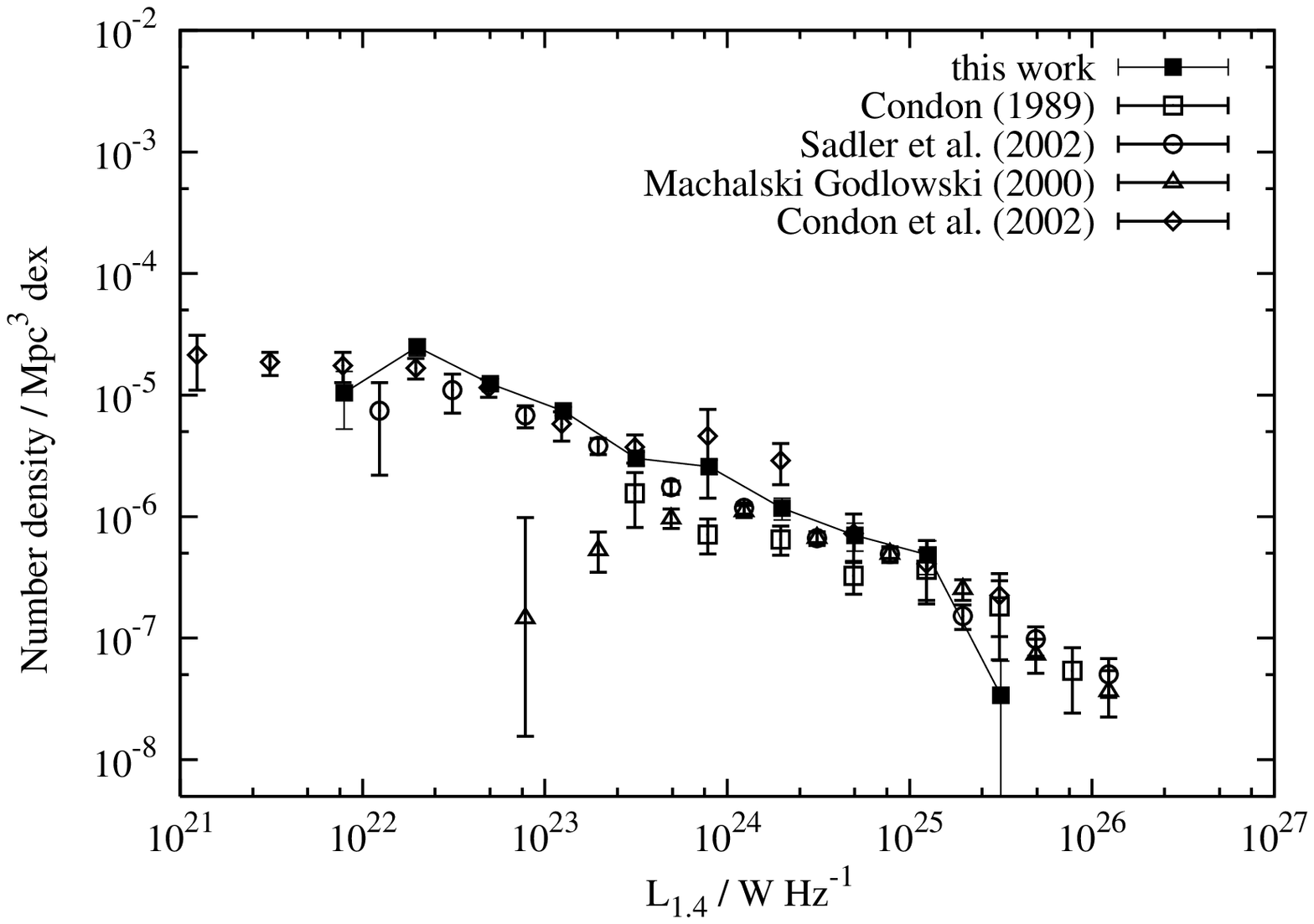}}
  \subfigure[Star-forming galaxies]{\includegraphics[height=0.4\textwidth,angle=270]{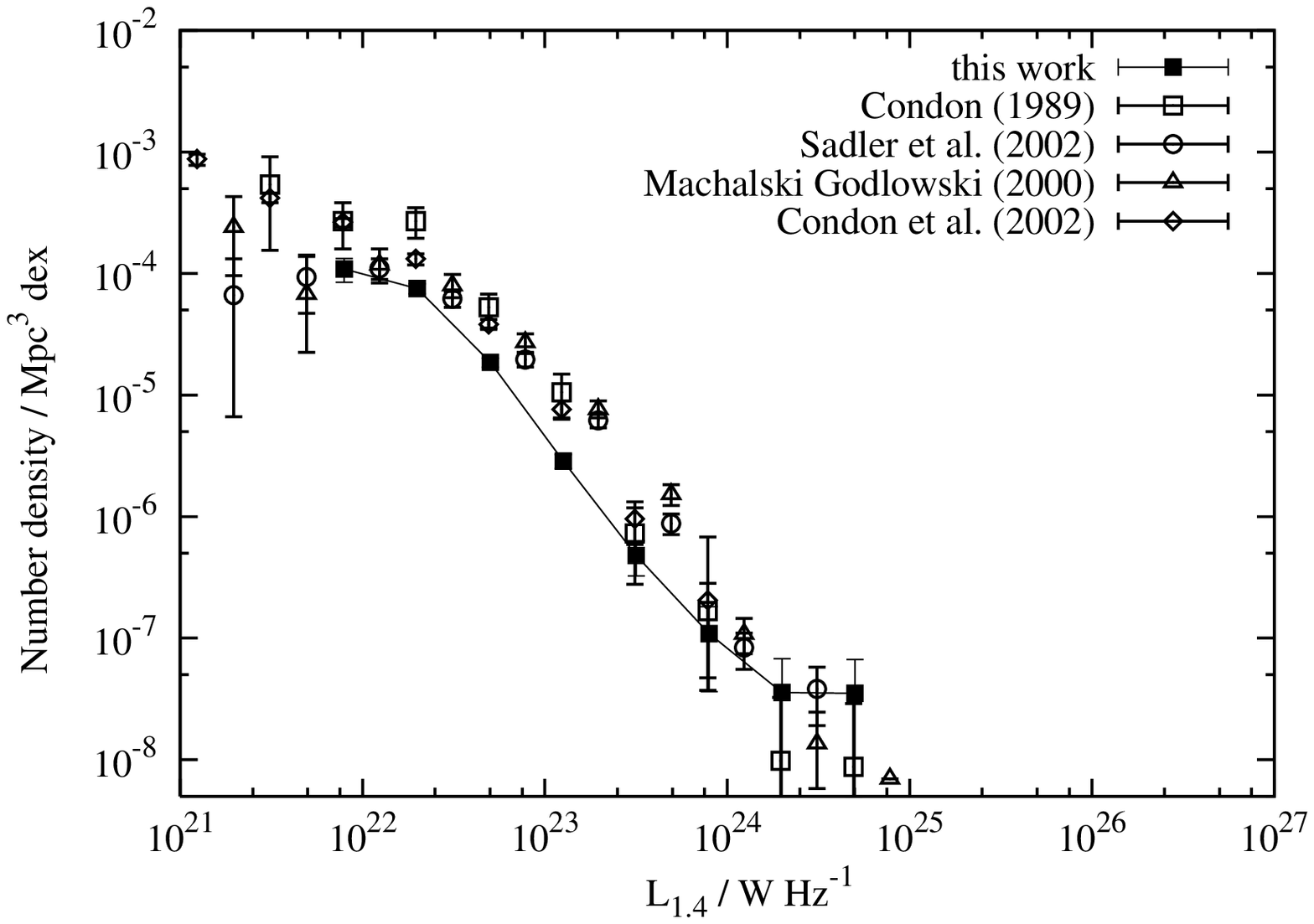}}
\caption{Local radio luminosity function at 1.4 GHz for {\itshape{(a)}} AGNs, and {\itshape{(b)}} star-forming galaxies as classified by a demarcation in the $\Dn\/(4000)$-$\frac{\LGHz}{\Mstar\/}$ plane \protect\cite{Paper1}. There is good agreement between our work and other studies for $\LGHz > 10^{23}$~\WHz\/. However, the luminosity functions diverge at luminosities below this value. This is especially evident in the case of the star-forming galaxies, and is due to our cutoff in r-band optical magnitude and the $\Mr\/$-$\LGHz$ correlation (see Figure~\ref{fig:opticalDropouts}).}   
\label{fig:RLFsplit}
\end{figure*}

The sample is split into star-forming galaxies and AGNs. Traditionally, this is done using the emission line properties of the objects as given by their location in the [O{\small{III}}]~5007/H$\beta$-[N{\small{II}}]~6583/H$\alpha$ plane \cite{BaldwinEA81,SadlerEA02,KauffmannEA03b}. However, Best \etal\/ \shortcite{Paper1} found that the radio and emission line AGN activity are independent phenomena, and argued that the 4000\AA\/-break, $\Dn\/(4000)$, represents the best way of separating the radio AGNs from their star-forming counterparts. This break is a consequence of accumulation of a large number of spectral lines in a narrow region in wavelength, and departs significantly from unity for older, metal rich galaxies \cite{KauffmannEA03b}. Using derived star-formation rates, Best \etal\/ (2005a; their Fig. 9) calculated the contribution of stellar emission to the 1.4 GHz luminosity and compared this with measured values to derive a demarcation in the $\Dn\/(4000)$-$\frac{\LGHz}{\Mstar\/}$ plane separating the objects into radio AGNs and star-forming galaxies. Since in this work we are primarily concerned with the radio AGN activity, we adopt this demarcation. Stellar mass $\Mstar\/$ for each object was evaluated by using the Petrosian z-band magnitude, and assuming that all stars are of solar type. Corrections for dust obscuration were made by comparing z and K-band fluxes. Nikolic~\shortcite{Nikolic05} found $z-K=2.5$ for our sample, and thus the stellar masses derived from z-band luminosities are underestimated by a factor of $3.2$. Proceeding in this way, we find 653 radio loud AGNs and 538 star-forming galaxies. The resultant RLFs are given in Table~\ref{tab:RLF} and plotted in Figure~\ref{fig:RLFsplit}.

\begin{figure}
\centering
  \includegraphics[height=0.48\textwidth,angle=270]{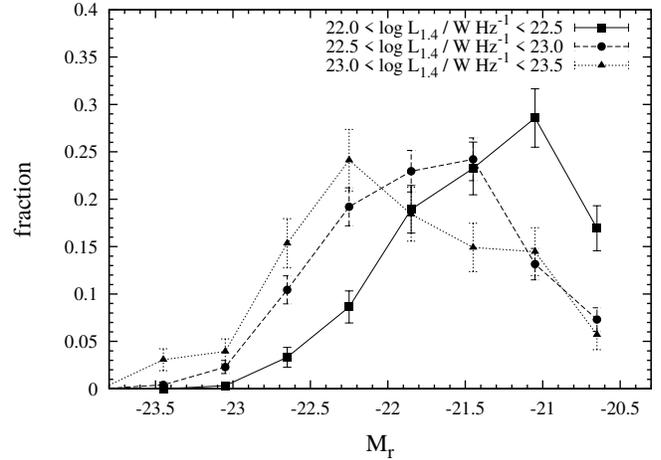}
\caption{Cuts through the bivariate radio-optical luminosity function. At high ($\LGHz >10^{23}$~\WHz) radio luminosities only very few objects have low enough r-band absolute magnitudes to be affected by our optical completeness limit. However, objects with optical magnitudes at or below our completeness cutoff of $\Mr\/=-20.45$ become important for $\LGHz <10^{23}$~\WHz\/. These missing sources account for the differences between our flux- and volume-limited luminosity function, and those derived from samples that are only flux-limited.}
\label{fig:opticalDropouts}
\end{figure}

The total luminosity function is in good agreement with previous studies at high luminosities, $\LGHz > 10^{23}$~\WHz\/. At luminosities below this value, however, our luminosity function is systematically lower than those of other authors. It is due to the fact that we are using a volume-limited sample, and have introduced a cutoff of $-20.45$ in r-band magnitude. The radio-optical correlation at low ($\LGHz < 10^{23}$~\WHz\/) luminosities means that in discarding objects with lower r-band magnitudes we are also excluding some radio sources. In Figure~\ref{fig:opticalDropouts} we illustrate this effect by binning our sample in radio luminosity, and plotting distributions of r-band magnitudes in each bin. There are relatively few objects with $\Mr\/ > -21.8$ at higher radio luminosities; however, when $\LGHz < 10^{23}$~\WHz\/ there is a significant contribution from galaxies with $\Mr\/>-21.8$, suggesting that there is also an appreciable number of low luminosity radio sources with $\Mr\/>-20.45$ that we have discarded due to our optical completeness limit. Not surprisingly, the difference between our luminosity function and those for flux-limited samples becomes apparent at around $10^{22.5}$~\WHz\/.

It is worth noting that the 4000\AA\/-break is not the only way to separate star-forming galaxies from radio AGNs in our sample. Another possible method is to use the concentration index, defined as the ratio of Petrosian 50-to-90\% light radii, $C = R_{50}/R_{90}$. Low values of $C$ ($< \sim 0.33$) correspond to early-type galaxies, which are more likely to harbour powerful radio sources, while late-type (star-forming) galaxies are those with $C>0.375$. We have carried out the subsequent analysis using both the $\Dn\/(4000)$ and concentration index diagnostics to split up our sample, and found very similar results. Therefore, only the 4000\AA\/-break demarcation findings are presented here.

\subsection{Bivariate luminosity function}
\label{sec:BLF}

\begin{figure}
  \centering
  \includegraphics[height=0.48\textwidth,angle=270]{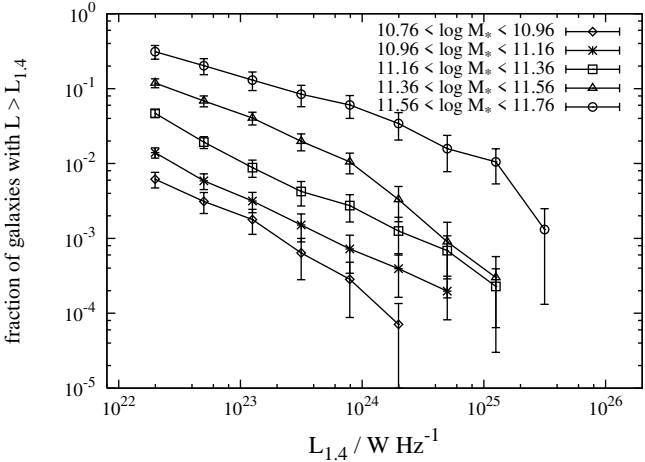}
  \caption{Bivariate luminosity function, with objects binned by stellar mass as calculated from the z-band magnitude (see text). Cumulative fraction of AGNs within a given $\Mstar\/$ range brighter than the specified radio luminosity ($fraction \left( L \right) = \frac{N_{\LGHz >{L}}}{N_{total}}$  is shown.}
  \label{fig:BLF}
\end{figure}

We construct a bivariate luminosity function by binning all objects in stellar mass and plotting the radio luminosity function for each bin. Fractions plotted represent the number of AGNs in a certain stellar mass range brighter than a given radio luminosity, divided by the total number of objects in that stellar mass bin. Again, corrections using the $V/\Vmax\/$ method \cite{Condon89} were applied. As r and z spectroscopic bands largely track the same population of old bulge stars, we expect completeness in r-band to imply that our sample is also complete in z-band, and hence in stellar mass. The bivariate luminosity function is shown in Figure~\ref{fig:BLF}.

The most striking feature of the bivariate luminosity function is the strong mass dependence of the radio loud fraction (equal to the cumulative fraction in the lowest radio luminosity bin). This result was also found by Best \etal\/ \shortcite{Paper2}, with the radio loud fraction $f_{RL} \propto \Mstar\/^{1.8}$. A vital question is whether this implies that in the more massive hosts the radio sources are on for longer, or whether they are simply triggered more frequently (with the duration of a typical active phase being the same for all masses). We address this question in subsequent sections by employing detailed radio source modeling.

\section{Radio source model}
\label{sec:radioSourceModel}

\subsection{Source evolution}
\label{sec:individualPDtracks}

We model radio loud AGNs as an evolutionary phase in the lifetime of every galaxy. The source is radio loud when the synchrotron jet is ``on''; and once it switches off or the source luminosity falls below the detection threshold, the source becomes radio quiet. We assume the black hole mass is already in place when the jet switches on, consistent with results of various semianalytic models (e.g. Kauffmann \& Haehnelt 2000; Bower \etal\/ 2006; Croton \etal\/ 2006) and observed black hole accretion rates in quasars \cite{HopkinsEA06,YuTremaine03}, and hence jet injection is represented by a top hat function, with the durations of on and off phases denoted as $\tOn\/$ and $\tOff\/$.

\begin{figure}
  \begin{center}
    \psfrag{rhoCore}{\large $\rho_{\rm core}$}
    \psfrag{betaCluster}{\large $\beta_{\rm cluster}$}
    \psfrag{betaGalaxy}{\large $\beta_{\rm galaxy}$}
    \psfrag{rCore}{\large $r_{\rm core}$}
    \psfrag{rTrans}{\large $r_{\rm trans}$}
    \psfrag{cocoon}{\large cocoon}
    \psfrag{bow shock}{\large bow shock}
    \psfrag{jet}{\large jet}
    \psfrag{black hole}{\large black hole}
    \psfrag{contact}{\large contact}
    \psfrag{discontinuity}{\large discontinuity}
    \psfrag{shocked gas}{\large shocked gas}
    \psfrag{hotspot}{\large hotspot}
    \includegraphics[height=0.35\textwidth]{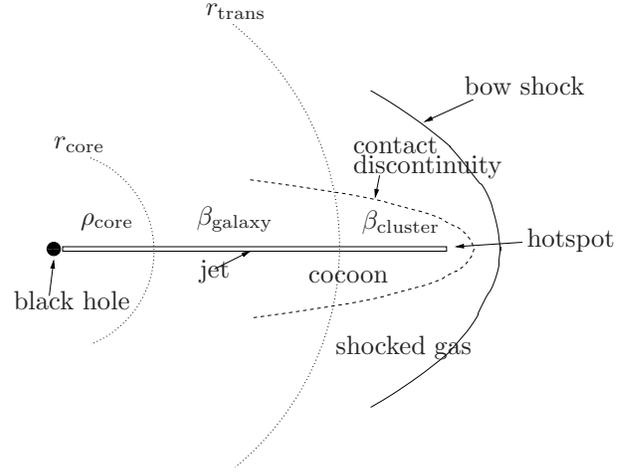}
  \end{center}
  \caption{Radio source and environmental parameters. The collimated jet expands into the ICM described by a core and double power-law, and eventually terminates in a hotspot, inflating a cocoon of synchrotron-emitting radio plasma. Because the jet is supersonic, a bow shock forms ahead of the cocoon, and a contact discontinuity separates the entrained gas from the radio plasma (see Kaiser \& Alexander 1997).}
  \label{fig:atmosphere}
\end{figure}

To account for the initial rise in the source radio power, we assume the source initially evolves in a flat atmosphere (such as a galaxy core), followed by two power-law profiles of the form $\rho (r) = \rhoCore\/ \left( \frac{r}{\rCore\/} \right)^{-\beta}$ corresponding to expansion within a galaxy, followed by a steeper cluster atmosphere (see Figure~\ref{fig:atmosphere}). We adopt the models of Alexander \shortcite{Alexander00} for the initial evolution within the core, and Kaiser \& Alexander \shortcite{KA97} and Kaiser \etal\/ \shortcite{KDA97} for evolution in a power-law profile. As the radio source ages, it will suffer adiabatic, synchrotron and inverse Compton losses. Once energy supply from the jet ceases, we assume the radio luminosity quickly drops to a value below our detection threshold. In practice, the cocoon will enter a ``coasting'' phase \cite{KaiserCotter02}; however the radio luminosity declines rapidly in this phase, rendering our approach sufficiently accurate.

These models describe the evolution of powerful (FR-II) radio sources. Although our sample consists of both these objects and the less powerful FR-Is, our approach is justified in the following sections by the relative paucity of resolved sources with FR-I morphologies.

\begin{figure}
\centering
	\includegraphics[height=0.48\textwidth,angle=270]{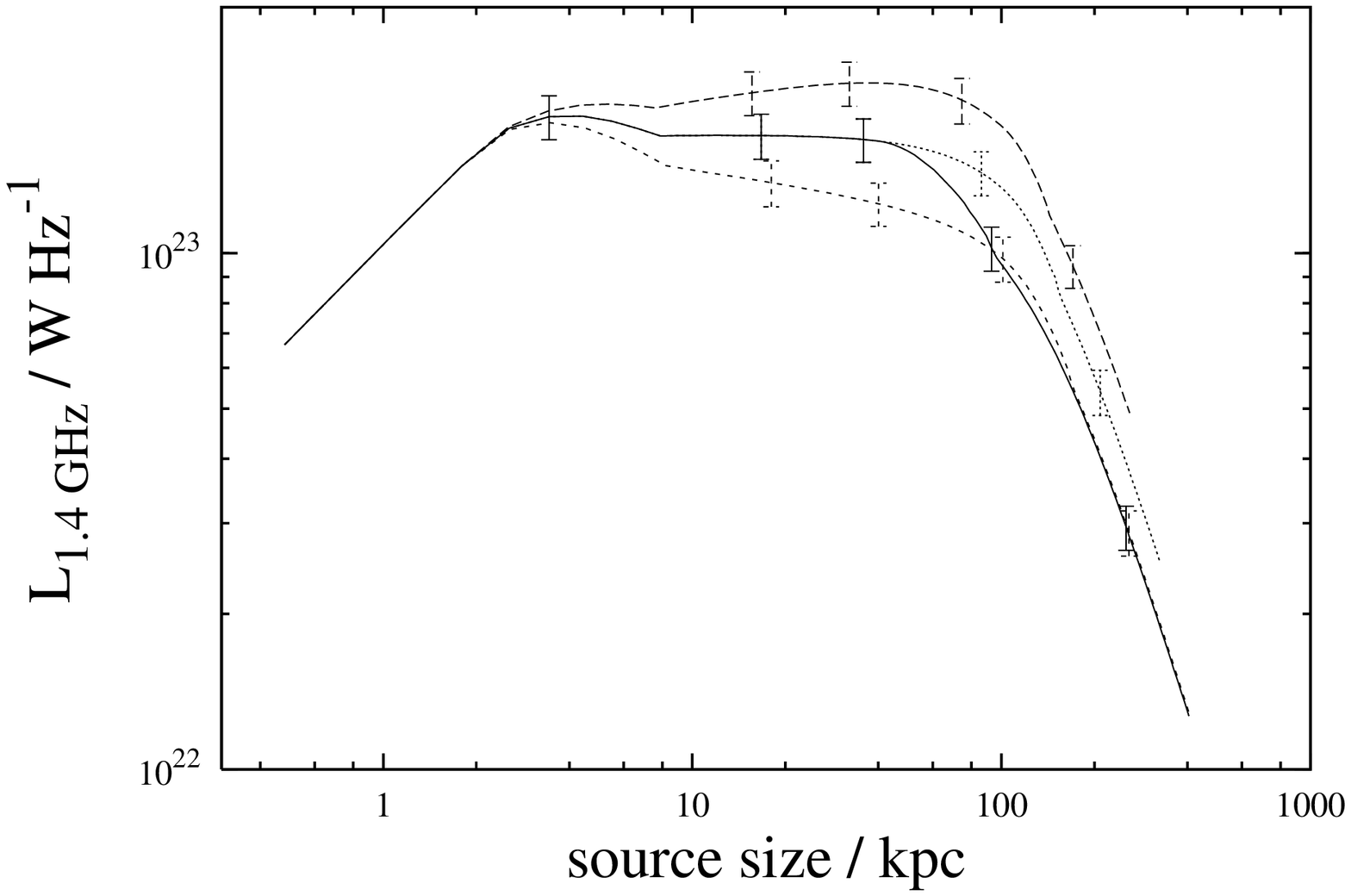}
\caption{Example radio luminosity$-$size tracks for sources evolving in different atmospheres. In all cases the adopted parameters are $\rhoCore\/=3.7 \times 10^{-22}$~\kgcm\/ (corresponding to $\nCore\/=0.2$~\cc\/), $\rCore\/=1$~kpc, $\betaCluster\/=1.9$, $\RT\/=2$ and $\Qjet\/=3 \times 10^{35}$~W. Evolution for $\tOn\/ = 5 \times 10^8$~years is shown. Varied parameters are the inner density exponent, $\betaGalaxy\/$, and transition radius $\rTrans\/$. Long-dashed curve is for $\betaGalaxy\/=0.8$, $\rTrans\/=50$~kpc; dotted curve is for $\betaGalaxy\/=1.0$, $\rTrans\/=50$~kpc; short-dashed curve is for $\betaGalaxy\/=1.2$, $\rTrans\/=50$~kpc; and solid curve for $\betaGalaxy\/=1.0$, $\rTrans\/=20$~kpc. Also shown are the time markers of $10^6$, $10^7$, $3 \times 10^7$, $10^8$ and $3 \times 10^8$~years. The source lumininosity does not evolve substantially until it begins to suffer significant inverse Compton losses, which happens around the same time as the source enters the steeper part of the atmosphere.}
\label{fig:exampleTracks}
\end{figure}

For an active jet, source size and radio luminosity are functions of core density $\rhoCore\/$, core radius $\rCore\/$, transition radius $\rTrans\/$ between the galaxy and cluster power laws, density exponents $\betaGalaxy\/$ and $\betaCluster\/$, jet opening angle $\theta$ (related to the axial ratio $\RT\/$ of the source) and jet power $\Qjet\/$. In adopting values for these parameters we are guided by observations of nearby X-ray luminous elliptical galaxies and relaxed clusters. In the inner regions \cite{AllenEA06}, power law exponents $\beta \sim 0.8$-$1.2$ and core radius $\rCore\/ \sim 1$~kpc provide a good fit to $r \sim 10$~kpc. Density profiles of  nearby relaxed galaxy clusters \cite{VikhlininEA06} are well-fitted by a double power-law, with inner regions having $\betaGalaxy\/ \sim 0.8$-$1.1$, and $\betaCluster\/ \sim 1.8$-$2.6$. Transition between the two power laws occurs at $\rTrans\/ \sim 50$-$200$~kpc. As Figure~\ref{fig:exampleTracks} illustrates, radio source tracks in the power-size (P-D) plane are rather flat and hence not very sensitive to the exact value of $\betaGalaxy\/$ for $0.8 \leq \betaGalaxy\/ \leq 1.2$ until the late phases ($1-$~a few times~$10^8$~years) of their lifetimes, when the sources begin to suffer appreciably from inverse Compton losses \cite{KDA97} and/or enter the steeper part of the atmosphere. At that point, the losses will typically be the dominant factor in determining the radio power of the source. This allows us to fix  $\betaGalaxy\/=1.0$, $\betaCluster\/=1.9$ and $\rTrans\/=50$~kpc in our models. Conservation of mass then sets the electron density within the core to be $\nCore\/ \sim 0.1-$~fewtimes~$0.1$~\cc\/, and we adopt $\nCore\/ = 0.2$~\cc\/ from the Allen \etal\/ \shortcite{AllenEA06} sample. Following observations of Cygnus~A \cite{BegelmanCioffi89} we also set $\RT\/ = 2.0$, corresponding to a jet opening angle of $31^\circ$ \cite{KA97}.

\subsection{Dependence of predicted RLF on model parameters}
\label{sec:parameterSpace}

\begin{figure*}
\centering
	\subfigure[$\rTrans\/$]{\includegraphics[height=0.32\textwidth,angle=270]{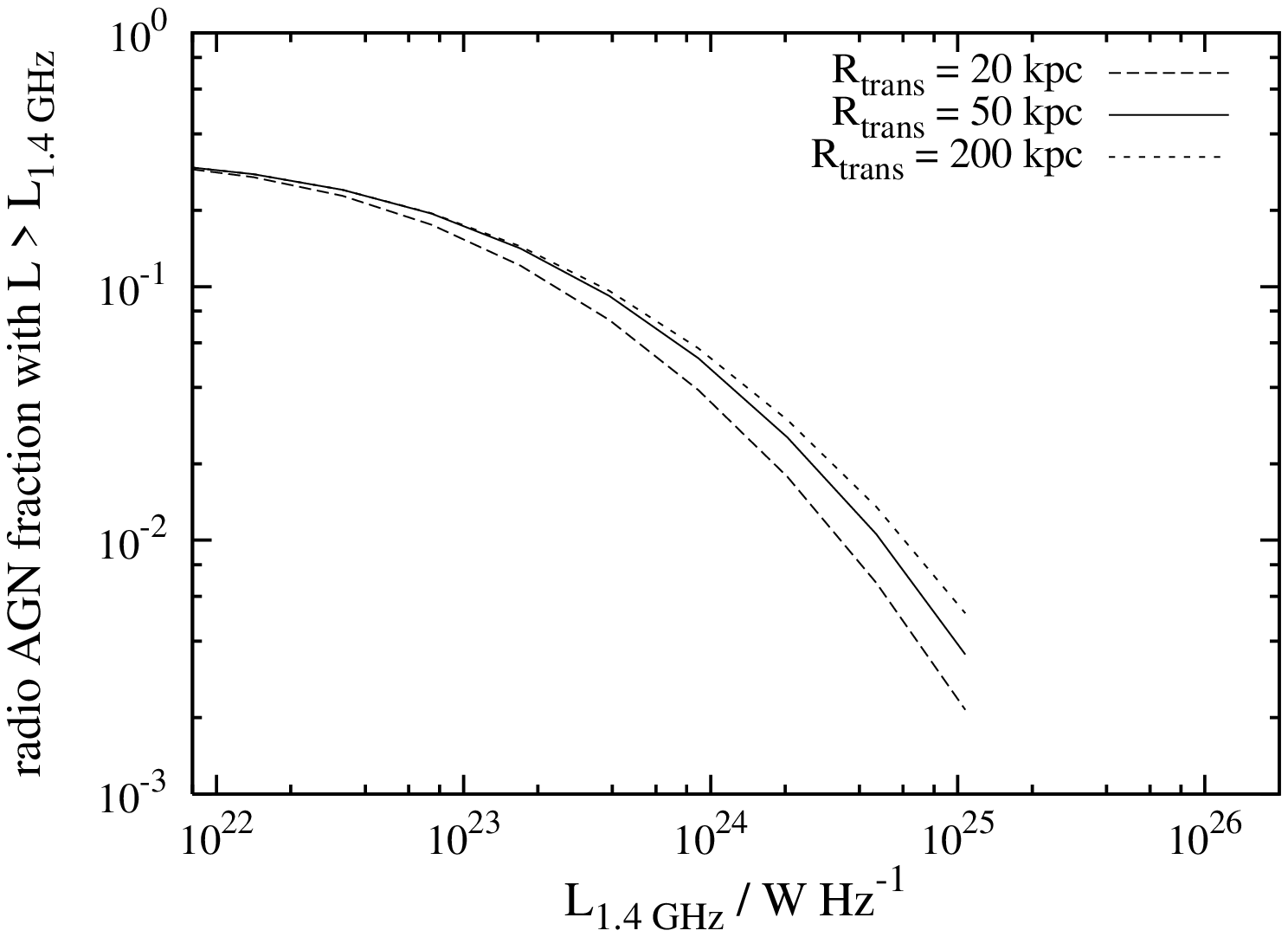}}
	\subfigure[$\betaCluster\/$]{\includegraphics[height=0.32\textwidth,angle=270]{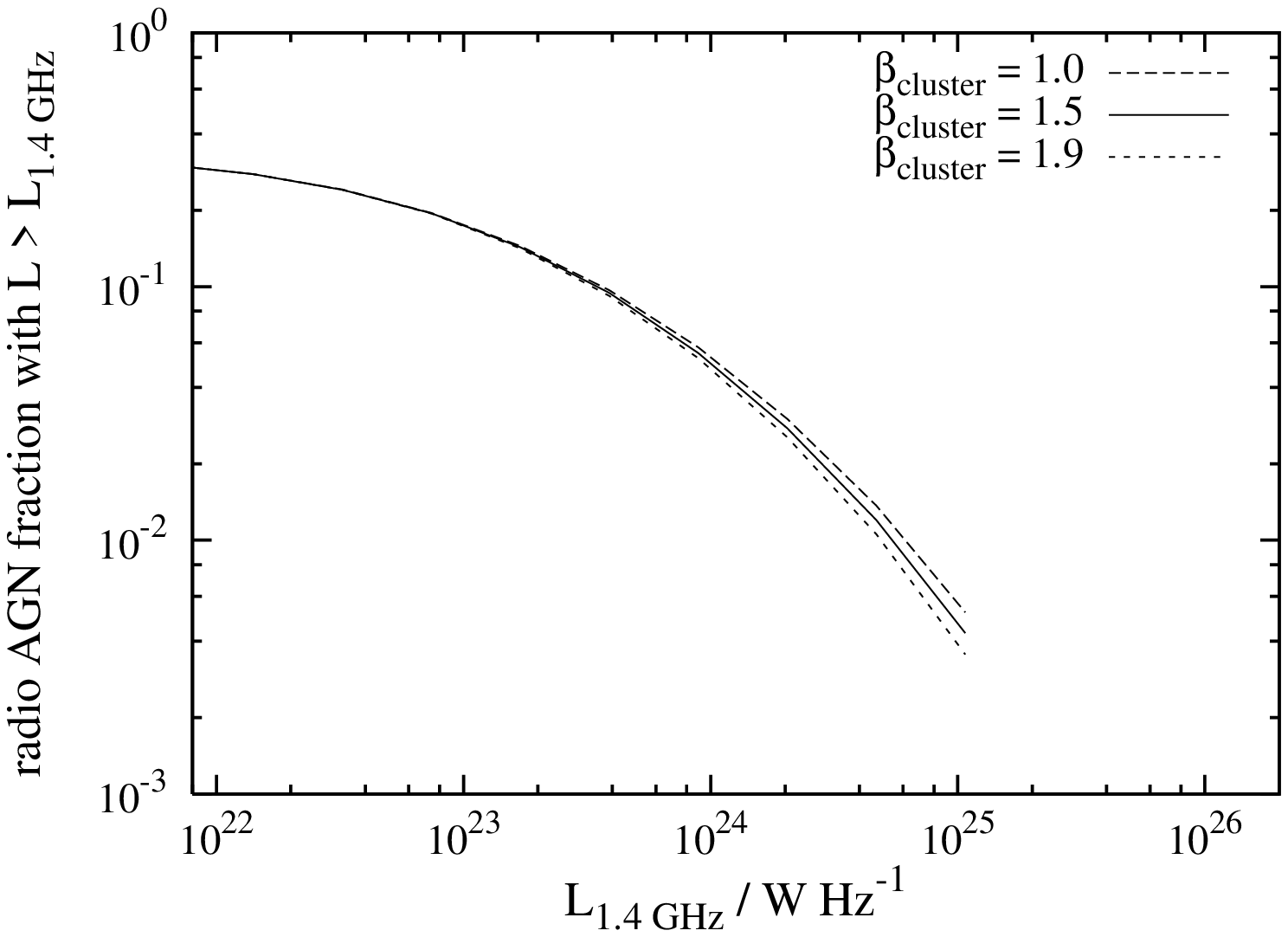}}
	\subfigure[$\tOn\/$]{\includegraphics[height=0.32\textwidth,angle=270]{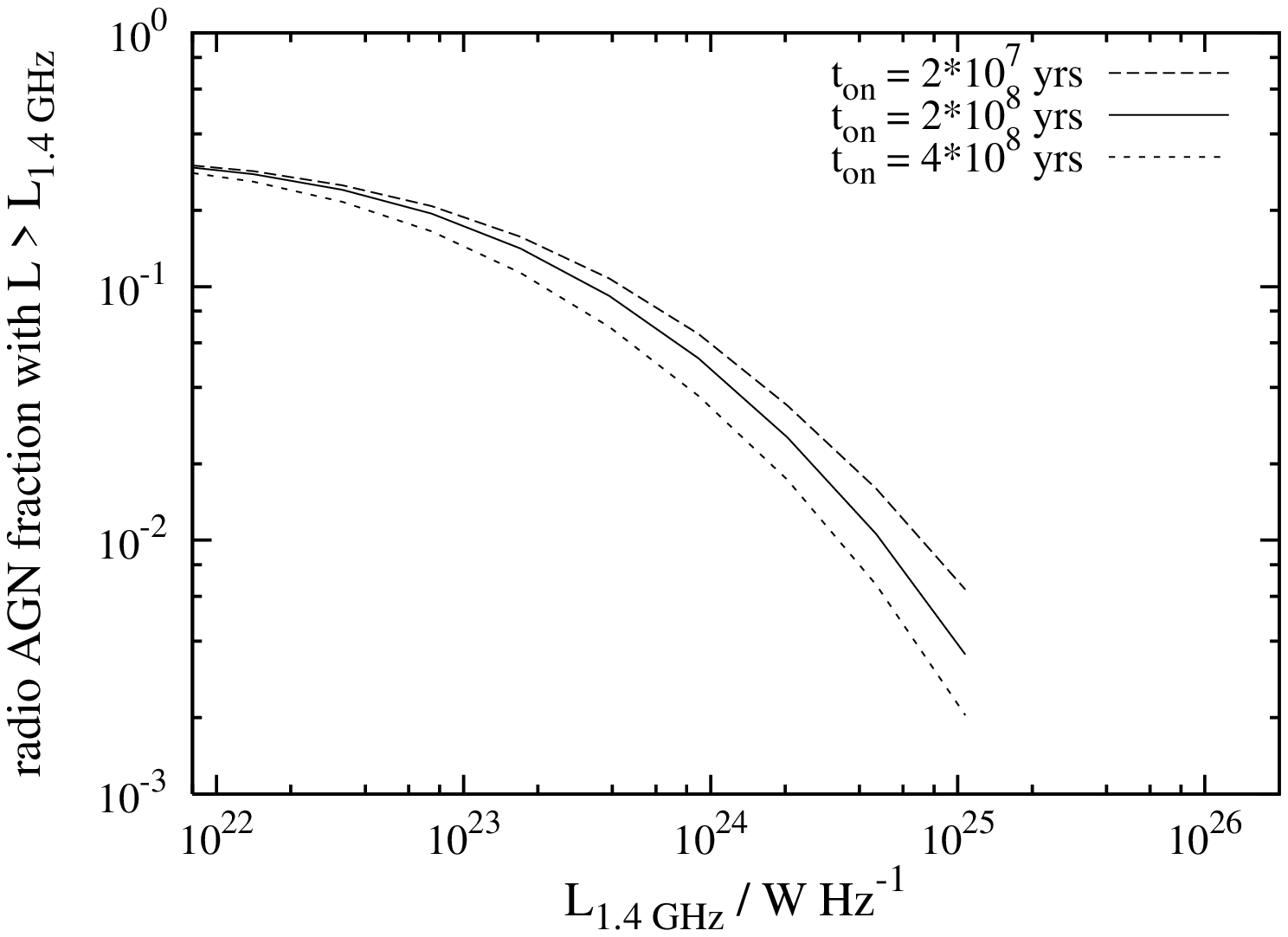}}
	\subfigure[$\frac{\tOff\/}{\tOn\/}$]{\includegraphics[height=0.32\textwidth,angle=270]{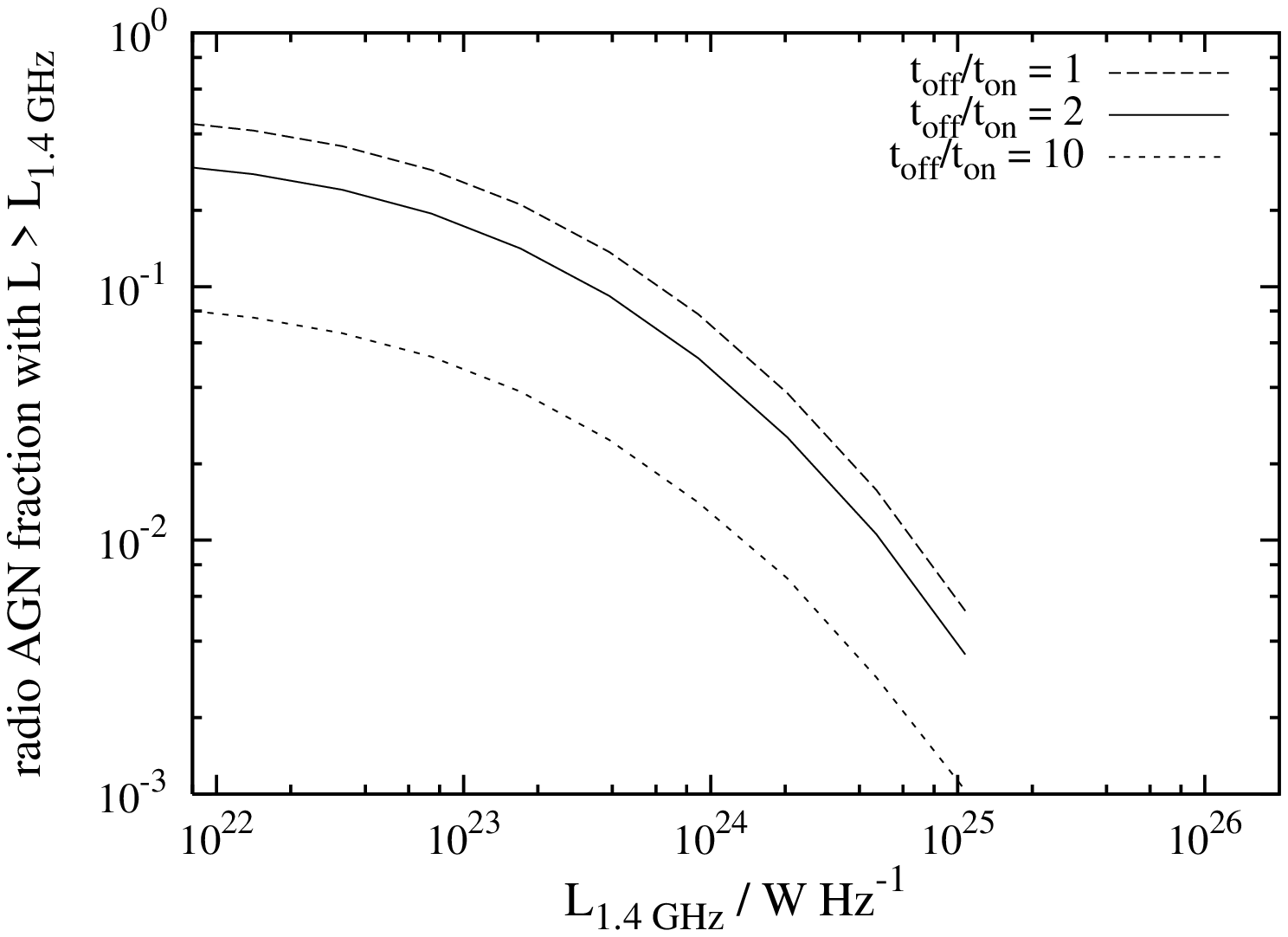}}
	\subfigure[$\betaGalaxy\/$]{\includegraphics[height=0.32\textwidth,angle=270]{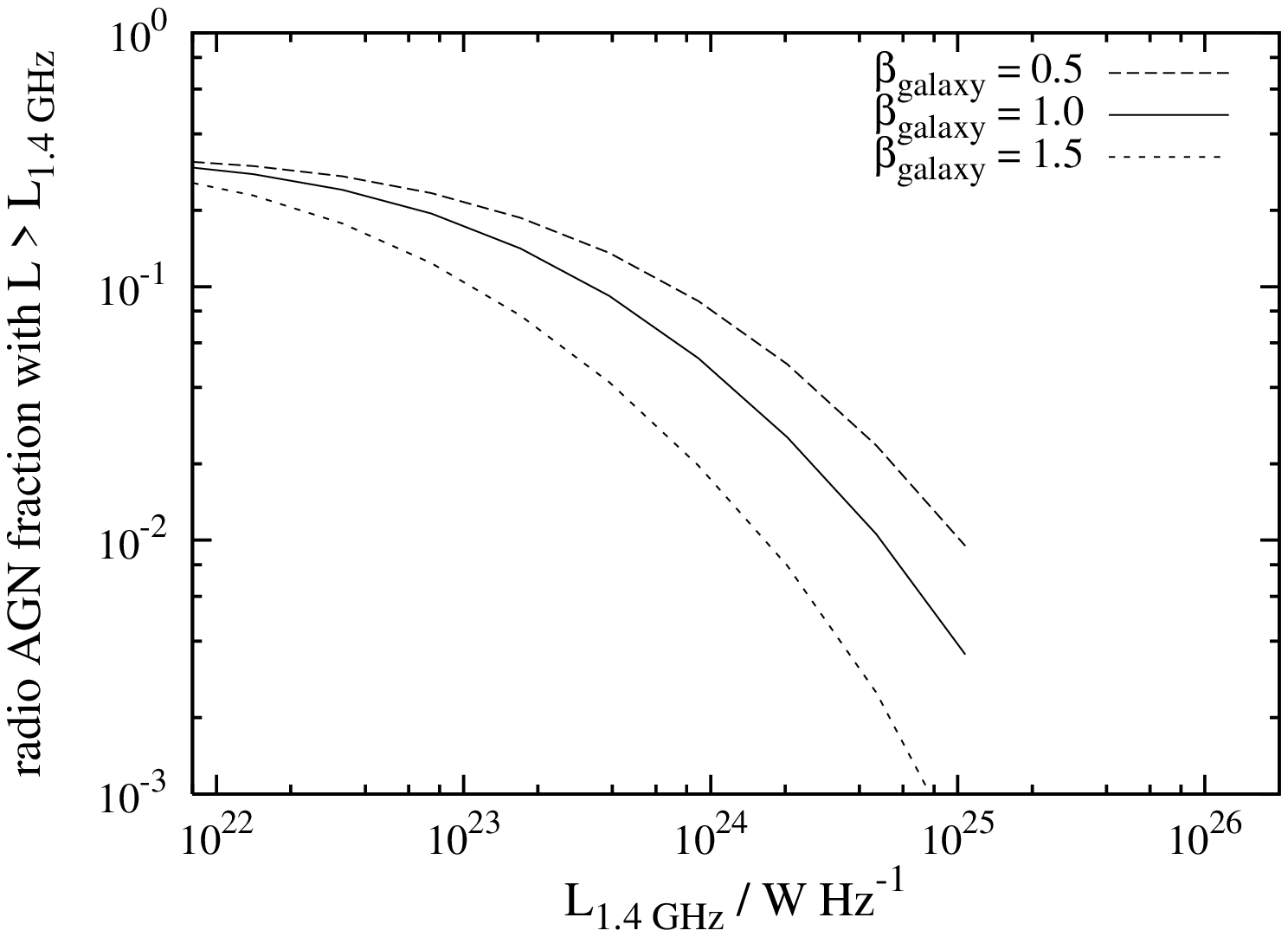}}
	\subfigure[$\bar{\Qjet\/}$]{\includegraphics[height=0.32\textwidth,angle=270]{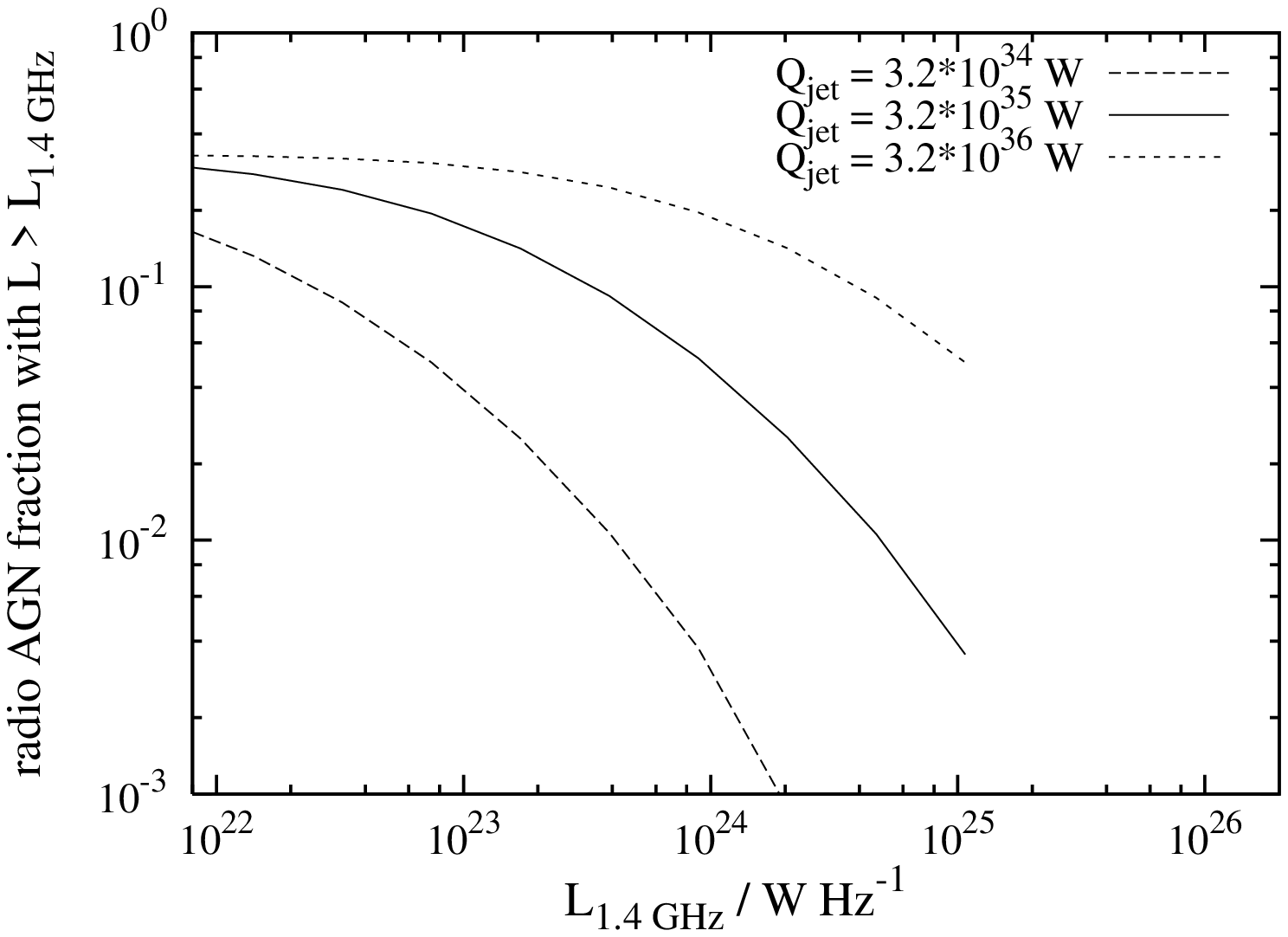}}
	\subfigure[$\rCore\/$]{\includegraphics[height=0.32\textwidth,angle=270]{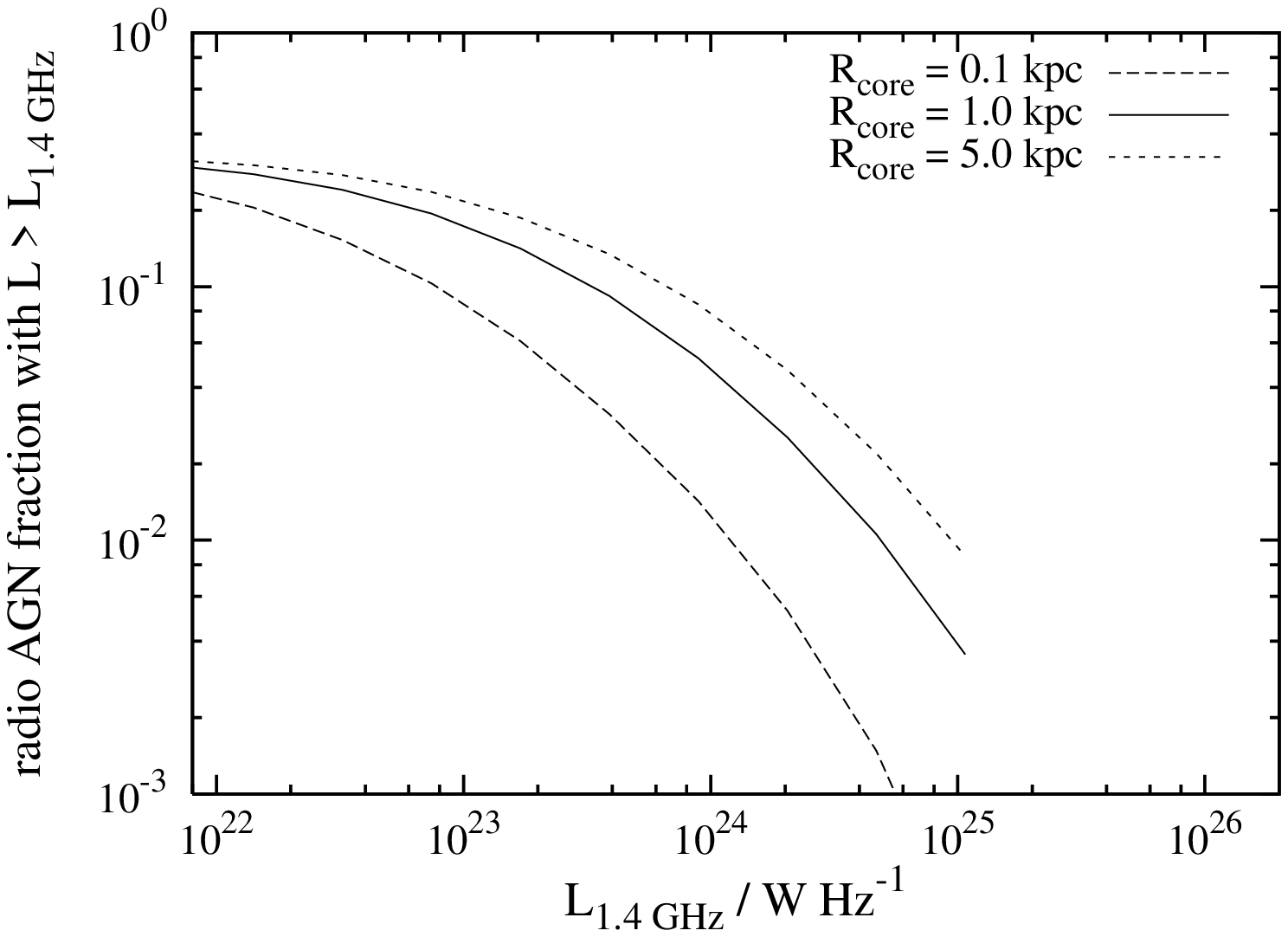}}
	\subfigure[$\nCore\/$]{\includegraphics[height=0.32\textwidth,angle=270]{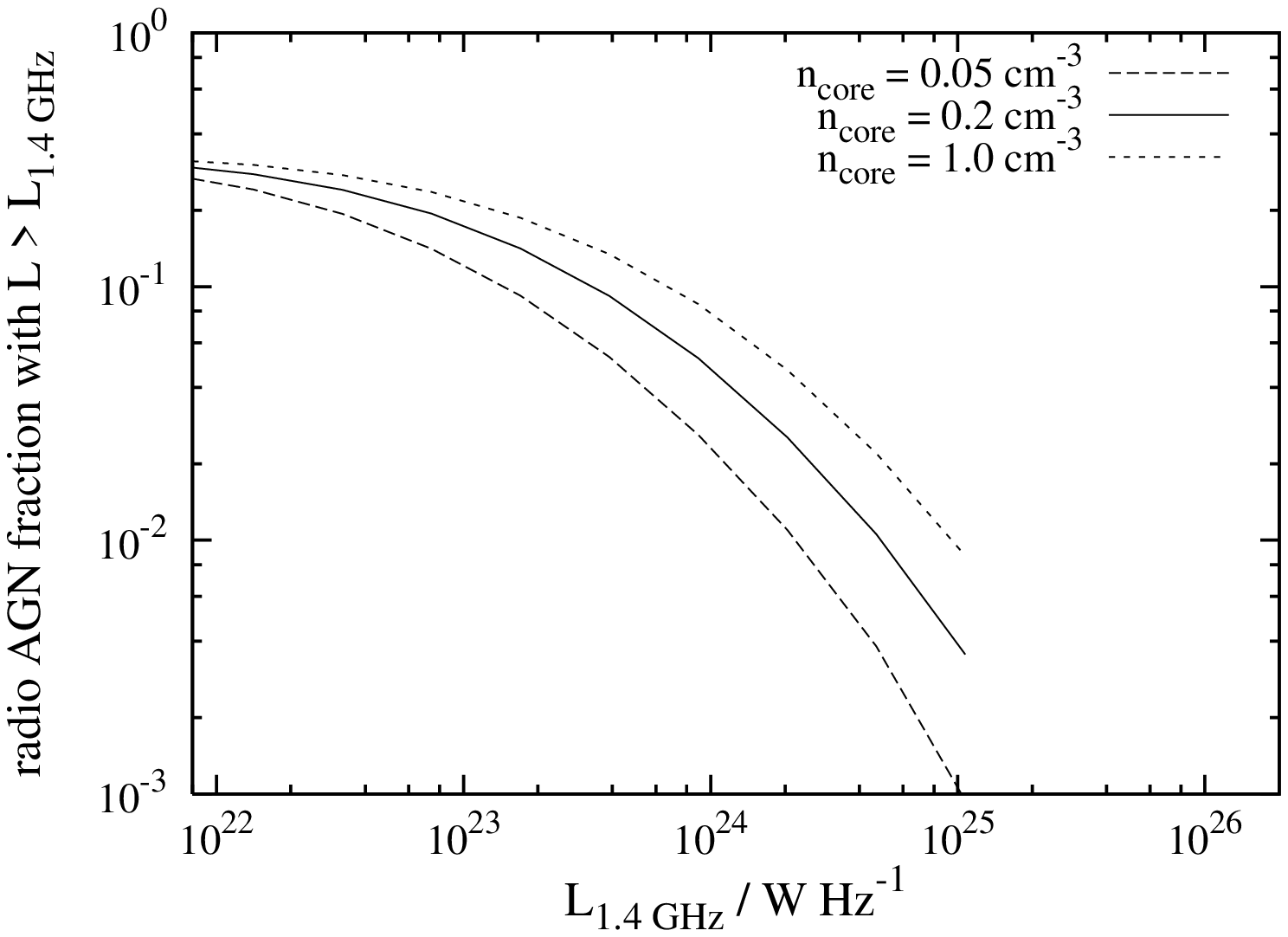}}
	\subfigure[$\RT\/$]{\includegraphics[height=0.32\textwidth,angle=270]{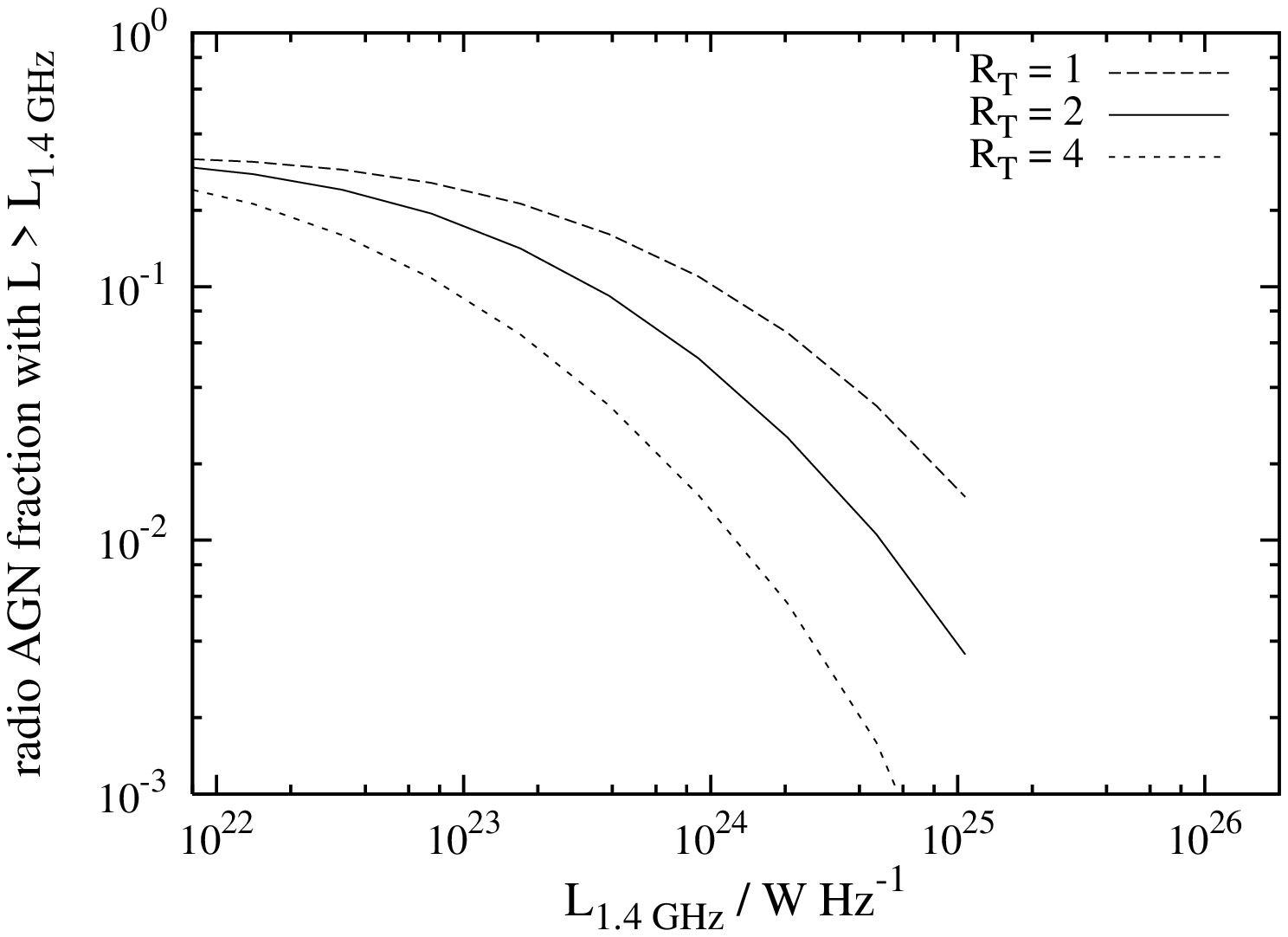}}
	\subfigure[$\sigmaQjet\/$]{\includegraphics[height=0.32\textwidth,angle=270]{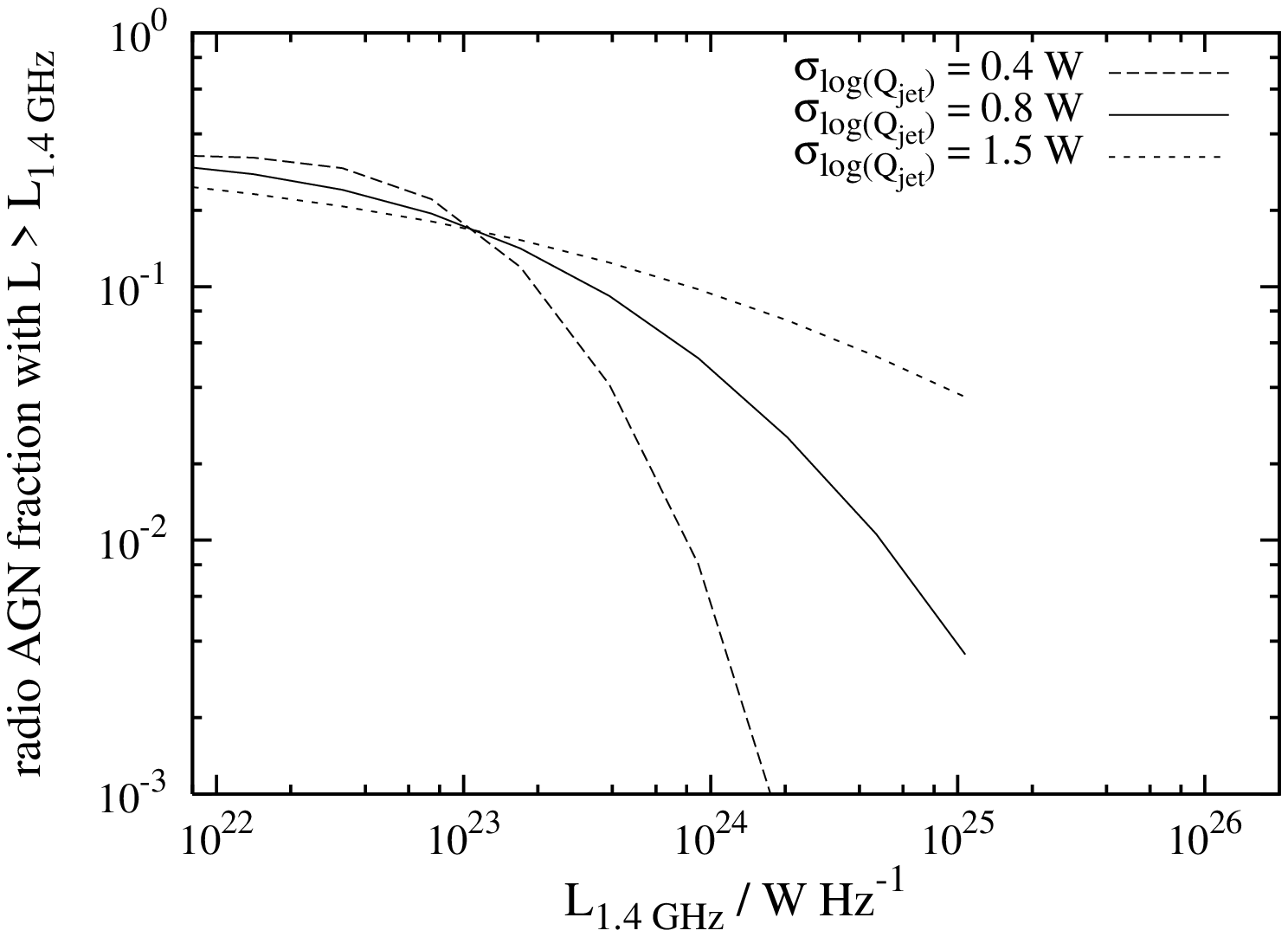}}
\caption{Dependence of the cumulative radio luminosity function (RLF) on input parameters. Base quantities used are: $\Qjet\/ = 3 \times 10^{35 \pm 0.8}$~W, $\nCore\/ = 0.2$~\cc, $\rCore\/ = 1.0$~kpc, $\betaGalaxy\/ = 1.0$, $\betaCluster\/ = 1.9$, $\rTrans\/ = 50$~kpc, cocoon axial ratio $\RT\/ = 2.0$, $\tOn\/ = 2 \times 10^8$~years and $\tOff\/ = 4 \times 10^8$~years. Transition radius $\rTrans\/$ between galaxy and cluster-type atmospheres ({\itshape{a}}), cluster power-law atmosphere exponent $\betaCluster\/$ ({\itshape{b}}), duration of on phase $\tOn\/$ ({\itshape{c}}), duty cycle $\tOn\//\tOff\/$ ({\itshape{d}}) and power-law exponent within the galaxy $\betaGalaxy\/$ ({\itshape{e}}) are not crucial to the shape of the RLF, with the notable exception of the duty cycle determining the number of radio quiet sources (see text for a discussion). The distribution of radio loud sources in luminosity is largely determined by mean jet power $\Qjet\/$ ({\itshape{f}}) and scatter ({\itshape{j}}) , core radius $\rCore\/$ ({\itshape{g}}), core density $\nCore\/$ ({\itshape{h}}), and the cocoon axial ratio $\RT\/$ ({\itshape{i}}).}
\label{fig:parameterSpace}
\end{figure*}

Given durations of jet on and off timescales, we can now predict the radio luminosity function (RLF) for a population of such sources observed at random stages in their evolution. Figure~\ref{fig:parameterSpace} explores the dependence of the RLF on input parameters. We adopt the same base quantities as above, in addition taking the scatter in jet power (as expected from the observed $\Mbh\/-\Mstar\/$ \cite{HaeringRix04,MagorrianEA98} and $\Qjet\/-\Mbh\/$ \cite{AllenEA06} relations) to be 0.8 dex. The jet is active for $\tOn\/ = 2 \times 10^8$~years, and off for time $\tOff\/ = 2 \tOn\/$.

To explore the sensitivity of our predicted cumulative RLFs on model parameters, each of these parameters is varied in turn (with the others fixed). The relative unimportance of the exact density profile of the atmosphere into which the source is expanding, as discussed above, is the reason very similar RLFs are predicted for various $\rTrans\/$ and $\betaCluster\/$ values in Figure~\ref{fig:parameterSpace}{\itshape{a}} and~{\itshape{b}}. Significantly, this implies that very similar RLFs are also predicted for sources of very different ages, so long as the duty cycle (i.e. $\tOn\//\tOff\/$) remains the same, and sources are not old enough for inverse Compton losses to dominate. This is shown in Figure~\ref{fig:parameterSpace}{\itshape{c}}, where we vary the jet on time, while keeping $\tOn\//\tOff\/$ constant. The duty cycle is the main factor determining the number of radio quiet sources, since the fraction of sources old enough to drop below the luminosity detection threshold is expected to be relatively small. This is illustrated in Figure~\ref{fig:parameterSpace}{\itshape{d}}, where duration of the active phase is kept constant, while the length of the quiescent phase is varied.

The flatness of the tracks for $t<$~few~Myrs for a range of power-law exponents in Figure~\ref{fig:exampleTracks} also implies that the RLF is also not very sensitive to changes in the density profile within the galaxy. This is seen in Figure~\ref{fig:parameterSpace}{\itshape{e}}. The ``characteristic'' luminosity of a source before it suffers significant inverse Compton losses is largely determined by mean jet power (Figure~\ref{fig:parameterSpace}{\itshape{f}}), core radius (Figure~\ref{fig:parameterSpace}{\itshape{g}}) and density (Figure~\ref{fig:parameterSpace}{\itshape{h}}), and axial ratio (Figure~\ref{fig:parameterSpace}{\itshape{i}}) of the source, in the sense that higher values of $\Qjet\/$, $\rCore\/$, $\rhoCore\/$ and lower value of $\RT\/$ result in higher characteristic luminosities. Finally, effects of scatter in jet power are shown in Figure~\ref{fig:parameterSpace}{\itshape{j}}. As expected, larger scatter in jet power results in less sources at the break (or ``characteristic'') luminosity, and hence a broader distribution of sources across the radio luminosity bins. The crucial feature of these plots is that they clearly show that changes in most of the parameters only result in redistributing the radio loud sources in luminosity. The only parameter that can significantly alter the predicted radio loud fraction is $\tOff\//\tOn\/$, i.e. the duty cycle. Hence observed radio loud fractions place tight constraints on relative durations of the radio active and quiescent phases in our sample.

Our sample is complete to $L_{\rm 1.4} = 8 \times 10^{22}$~\WHz\/, rather than the $10^{22}$~\WHz\/ plotted in Figure~\ref{fig:parameterSpace}. Therefore, only sources with luminosities brighter than this value can be used to constrain the models. Inspection of Figure~\ref{fig:parameterSpace} shows that the discussion above is still applicable, although jet and environmental parameters affect the total radio loud fraction to a larger degree for this higher $L_{\rm 1.4}$ value.

\section{Duration of radio active and quiescent phases}
\label{sec:results}

\subsection{Source sizes}
\label{sec:sourceSizes}

It is clear from Section~\ref{sec:parameterSpace} that observed RLFs can be used to constrain the duty cycles (i.e. relative lengths of the radio active and quiescent phases), however it is difficult to place constraints on actual values of $\tOn\/$ and $\tOff\/$. Instead, only tentative upper or lower limits can be placed on these. The former would correspond to a case where the best fit to the observed RLF is for a population of sources that do not suffer significant inverse Compton losses; and the latter to a case when they do.

A much better constraint on these timescales is obtained by considering source sizes. For a constant jet power, dynamical models of Kaiser \& Alexander \shortcite{KA97}, Kaiser \etal\/ \shortcite{KDA97} and Alexander~\shortcite{Alexander00} predict source sizes as well as radio luminosities as a function of time (see Figure~\ref{fig:exampleTracks}). Sources are observed at various stages in their lifetimes, and it is the oldest sources that give an estimate of either the duration of the active phase, or the age of the source when its radio luminosity falls below our detection threshold. The discussion of Section~\ref{sec:individualPDtracks} suggests the latter case will only occur for very weak jets, and even then this is unlikely.

Contour maps of the 1191 radio loud sources in the present sample allow source sizes to be determined. Sizes of sources with prominent lobes (FR-IIs) were determined manually. Defining source size for low luminosity FR-Is is not easy, since the source surface brightness decreases smoothly with distance from the nucleus. For the purposes of a comparative analysis within the presented sample, however, it is sufficient to adopt catalogued NVSS and FIRST major and minor axes FWHM.

Strictly speaking, the radio source model employed in this work only describes the evolution of FR-II sources. However, two points justify its applicability to the whole radio source population in our sample. Figure~\ref{fig:observedSizesPD} shows that there are relatively few bona fide FR-Is. The largest sources are invariably FR-IIs, and at smaller sizes the sample is dominated by sources that are only just resolved (and hence are those sources for which morphologies cannot be reliably determined). As all radio sources undergo an initial phase of supersonic expansion, we expect the FR-II model to provide an accurate description of this sub-population. Once the supersonic jets are disrupted, observed distributions of source sizes \cite{KaiserBest07} and theoretical considerations \cite{Alexander00} suggest source luminosities quickly drop below the detection threshold. We performed the analysis outlined below on both the full radio source sample, and with FR-Is excluded, and obtained identical results. This justifies our application of the radio source model to FR-I sources, and in what follows we present the results for the whole population.

\subsection{Orientation effects}
\label{sec:orientation}

No spectral information is available for the catalogued radio sources. This means various orientation effects must be considered when interpreting source properties such as size and luminosity. The observed quantities will only correspond to actual source parameters for sources that are observed face-on. Any other orientation (i.e. beaming) conspires to decrease the apparent source size, and increase its luminosity. In other words, the source must be moved right and down in the P-D plane from the observed position.

Consider a source subtending angle $\phi$ to the line-of-sight. If the source has size $D$, its projected size will be $D_{\rm app} = D {\rm sin} \phi$. The apparent luminosity is related to intrinsic luminosity via \cite{BlandfordKonigl79} $L_{\rm app} = L \delta_{\rm D}^{3+\alpha}$, where $\delta_{\rm D} = \left[ \left( 1-\beta_{\rm D}^2 \right)^{1/2} (1 - \beta_{\rm D} {\rm cos} \phi) \right]^{-1}$ is the Doppler factor for $\beta_{\rm D} = v_{\rm source}/c$ and $\alpha$ is the spectral index of the source.

Observations of powerful radio galaxies \cite{HardcastleEA98} suggest $v_{\rm source} \sim 0.8 c$, and adopting $\alpha=0.8$ we have

\begin{eqnarray}
  \frac{L_{\rm app}}{L} & \approx & (1-0.8 {\rm cos} \phi)^{-3.8} \nonumber\\
  \frac{D_{\rm app}}{D} & = & {\rm sin} \phi .
\label{eqn:beaming}
\end{eqnarray}

\begin{figure}
  \begin{center}
    \includegraphics[height=0.48\textwidth,angle=270]{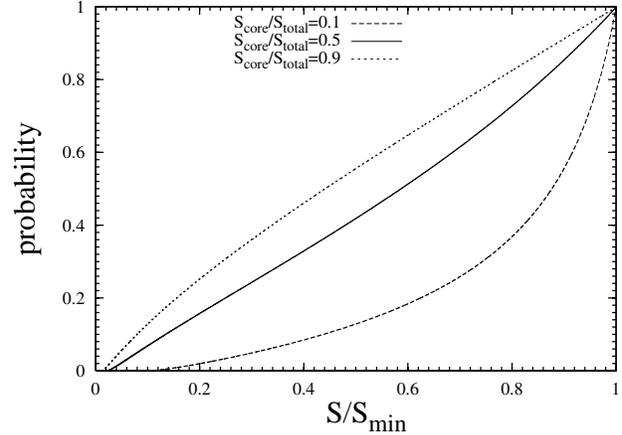}
  \end{center}
  \caption{Probability of detecting a source with flux below the catalogue threshold, $S<S_{\rm min}$. The non-relativistic extended component of the flux is not affected by beaming.}
  \label{fig:beaming}
\end{figure}

The solid angle subtended by $\phi$ is ${\tiny \Delta} \Omega = 2 \pi (1- {\rm cos} \phi)$. In accordance with AGN unification models there should be no preferred direction for the beaming, and hence this solid angle is directly related to the probability of the major axis of the radio source being directed within $\phi$ of the line-of-sight. A convolution of this probability with the expression for $L_{\rm app}$ in Equation~\ref{eqn:beaming} then gives the probability of a source with luminosity $L$ (or flux $S$) below the flux threshold (3.4~mJy for the present sample) being detected due to the beaming. Figure~\ref{fig:beaming} plots this probability as a function of intrinsic source flux.

Most radio sources contain a core component as well as more extended emission such as lobes. Since only the relativistic core component will be beamed, the level of sample contamination by low luminosity sources is sensitive to the adopted core-to-total flux ratio, with core-dominated sources prone to substantial beaming. In general this effect is difficult to quantify. However, the presented sample is volume-limited. This means that the objects were selected according to their optical properties, which are unaffected by orientation. Moreover, the correlation between optical and radio luminosities at the low luminosity end of the sample (Section~\ref{sec:observedRLF} and Figure~\ref{fig:opticalDropouts}) suggests there are very few sources with uncharacteristically high radio luminosities. Thus the radio luminosities of objects in the sample are unlikely to be greatly affected by beaming.

The apparent source sizes will not be affected by the beaming by anywhere near the same amount as the luminosities. For ${\tiny \Delta} \Omega / 4 \pi = 0.01$ Equation~\ref{eqn:beaming} gives $D_{\rm app} / D = 7$, while $L_{\rm app} / L_{\rm extended}=1/200$. For an appreciable solid angle of ${\tiny \Delta} \Omega / 4 \pi = 0.3$ this value drops to $D_{\rm app} / D = 1.4$, while $L_{\rm app} / L_{\rm extended}=1/25$. Hence any orientation-related corrections are likely to affect the derived radio luminosities to a much greater extent than source sizes. As a zeroth-order approximation, the source will move (almost vertically) down in the P-D plane. This introduces larger errors in the derived jet powers than source ages.

\section{Derived source properties}
\label{sec:derivedProperties}

\subsection{Jet powers}
\label{sec:jetPowers}

\begin{figure}
  \centering
  \psfrag{xLabel}{$M_\star / M_\odot$}
  \psfrag{yLabel}{${\rm log} \Qjet\/$ / W}
  \includegraphics[height=0.48\textwidth,angle=270]{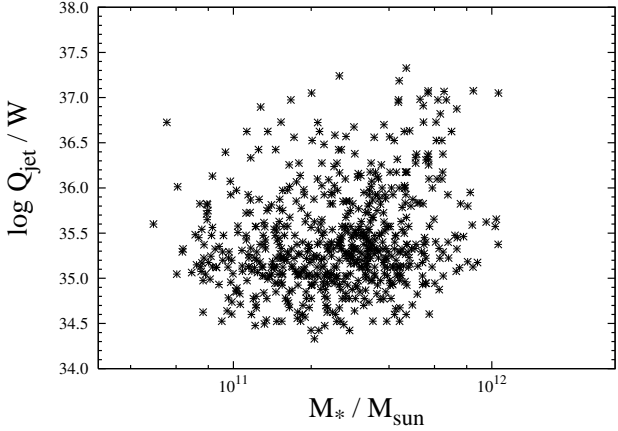}
  \caption{Distribution of jet power as a function of bulge mass. These are derived by combining observed source sizes and luminosities with the radio source model. Radio source and environmental parameters of Section~\ref{sec:individualPDtracks} are used for the modeling.}
  \label{fig:jetPowers}
\end{figure}

For a given atmosphere and cocoon axial ratio, a source with specified radio luminosity and size can only be described by a single evolutionary track. Hence, adopting the density profiles of Section~\ref{sec:parameterSpace} and once again taking $\RT\/=2$, the position of a source in the P-D plane uniquely defines its age and jet power. Proceeding in this fashion, we derived jet powers and ages for every source classified as a radio loud AGN. The corresponding jet powers are plotted as a function of stellar mass in Figure~\ref{fig:jetPowers}. Although the scatter is large, the more massive hosts contain less sources with lower ($\log \Qjet\/<35.5$~W) jet powers.

\begin{figure}
  \centering
  \psfrag{xLabel}{fractile}
  \psfrag{yLabel}{$\log (\Qjet\/$ / W$)$}
  \includegraphics[height=0.48\textwidth,angle=270]{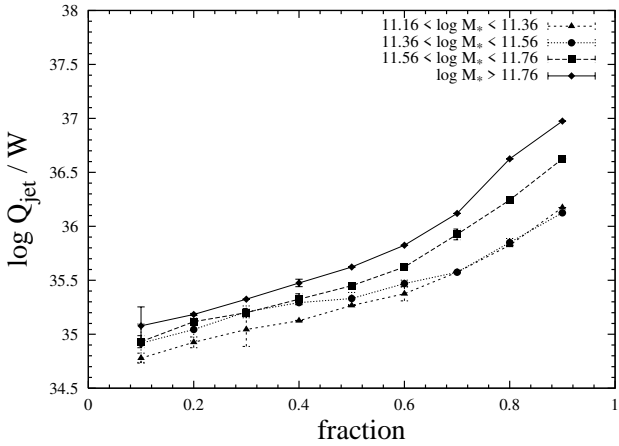}
  \caption{Fractile distribution of jet power for the top four stellar mass bins. Lowest mass bins are not shown due to large numbers of unresolved sources in these affecting the statistics significantly. Massive galaxies host more powerful jets.}
  \label{fig:fractileQjet}
\end{figure}

This point is illustrated in Figure~\ref{fig:fractileQjet}, where we plot the jet power fractile ranges for various stellar mass bins. In the bulk of the sample (0.3 to 0.8$-$fractile ranges) all mass bins follow a similar trend, with the more massive hosts consistently hosting more powerful radio sources.

The mean jet power (Table~\ref{tab:derivedParameters}) increases with stellar mass as $\bar{Q}_{\rm jet} \propto \Mstar\/^{0.8 \pm 0.2}$. Benson \etal\/ (2007) analysed a sample of $\sim 9000$ SDSS galaxies and found that almost all stellar mass is located in the spheroid component for the stellar masses considered in the present sample. Hence $\bar{Q}_{\rm jet} \propto M_{\rm bulge}^{0.8 \pm 0.2}$, or $\bar{Q}_{\rm jet} \propto \Mbh\/^{0.7 \pm 0.5}$ using the 0.3~dex scatter in the $\Mbh\/-M_{\rm bulge}$ relation \cite{HaeringRix04}. This is significantly less steep than the $\Qjet\/ \propto \Mbh\/^2$ expected from Bondi accretion, suggesting there exists some factor limiting the accretion/outflow process. The more massive galaxies also show a larger fraction of high $\Qjet\/$ values (i.e. a larger scatter at the high-$\Qjet\/$ end; see Table~\ref{tab:derivedParameters}). Since it is the number of powerful sources that determines the shape of the radio luminosity function at the bright, low number counts end, these values for the scatter in jet power are adopted when fitting the observed bivariate luminosity function below.

\begin{table*}
\begin{centering}
\begin{tabular}{c | ccccc} \hline
  $\log \Mstar\/ / M_{\odot}$ & $\log \bar{Q}_{\mbox{jet}}/W$ & $\sigmaQjet\/ /W$ & $t_{\rm on, median}$ / yr & $\frac{\tOff\/}{\tOn\/}$ & $\tOff\/$/yr  \\ \hline
  $> 11.76$     & $36.0$ & $1.0$ & $4 \times 10^6$  & 2   & $2 \times 10^7$ \\
  $11.56-11.76$ & $35.7$ & $0.7$ & $3 \times 10^6$  & 6   & $4 \times 10^7$ \\
  $11.36-11.56$ & $35.6$ & $0.7$ & $2 \times 10^6$  & 20  & $8 \times 10^7$ \\
  $11.16-11.36$ & $35.6$ & $0.7$ & $8 \times 10^5$  & 50  & $8 \times 10^7$ \\
  $10.96-11.16$ & $35.5$ & $0.7$ & $-$              & 100 & $-$ \\ \hline
\end{tabular}
\caption{Jet powers and timescales determined from observed RLFs and source distribution in the P-D plane. Jet powers and median source ages are determined from observed source sizes using parameters of Section~\ref{sec:parameterSpace}. Quiescent phase duration is found from $\frac{\tOff\/}{\tOn\/}$ ratios obtained by fitting to the observed luminosity function for each $\Mstar\/$ bin.}
\label{tab:derivedParameters}
\end{centering}
\end{table*}

\subsection{Independence of optical and radio AGN activity}
\label{sec:radioOpticalIndependence}

Best \etal\/ (2005b) considered the relation between emission line and radio AGN activity in their sample in two ways. First, they looked at the distribution of [O{\small{III}}] luminosities as a function of radio luminosity, and found there to be no correlation between the two quantities for their (mostly low $\LGHz\/$) sample (their Fig. 5). Secondly, Best \etal\/ compared the radio loud AGN fractions as a function of stellar mass in emission line and optically inactive AGNs, finding no significant difference between the two (their Fig. 7). These findings suggest that whether or not a given object is classified as an emission line AGN or a normal galaxy has no bearing on how likely it is to host a radio AGN. An identical analysis of our sample confirmed these results.

\begin{figure}
\centering
  \includegraphics[height=0.48\textwidth,angle=270]{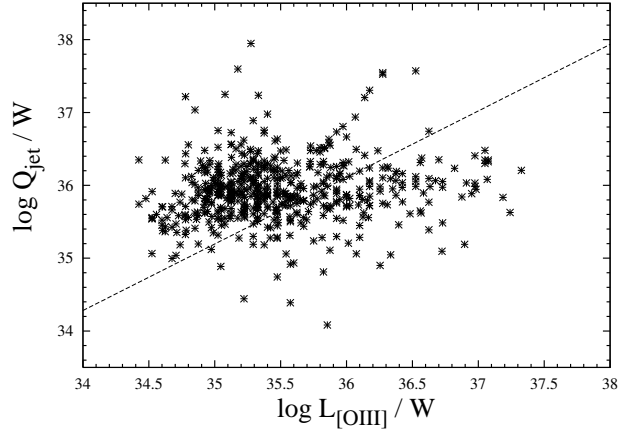}
\caption{Distribution of jet power with [O{\scriptsize{III}}] line luminosity for the radio AGN subsample. The Rawlings \& Saunders (1991) relation is shown by a dashed line.}
\label{fig:QjetVsLoiii}
\end{figure}

Using radio source models, we extend this analysis to investigation of the radio jet power dependence on [O{\small{III}}] line luminosity, as shown in Figure~\ref{fig:QjetVsLoiii}.

The [O{\small{III}}] luminosity probes the narrow-line region, and is thus indicative of the AGN bolometric luminosity. No correlation between $Q_{\rm jet}$ and $L_{\rm \small [O{\scriptsize III}]}$ is observed in the largely low radio luminosity sample presented here. By constrast, using spectral ages to estimate jet powers, Rawlings \& Saunders (1991) found a tight relation between narrow-line optical luminosity and jet power in powerful radio sources; this is shown in Figure~\ref{fig:QjetVsLoiii} by a dashed line. The fact that no such correlation exists for low luminosity radio sources confirms the claim of Best \etal\/ that radio and optical AGN activity are distinct phenomena in these objects. An intriguing possibility is that these correspond to different accretion states. In this picture, powerful radio jets would be fuelled when the accretion disk is in a radiatively inefficient, low luminosity state; while optical AGN are detected when the disk is radiatively efficient and correspondingly the jet power is low \cite{NarayanEA98,Meier01}.

\subsection{Active timescales}
\label{sec:tOn}

\begin{figure*}
\centering
	\subfigure[${\rm log} \Mstar\/ / M_{\odot} > 11.76$]{\includegraphics[height=0.48\textwidth,angle=270]{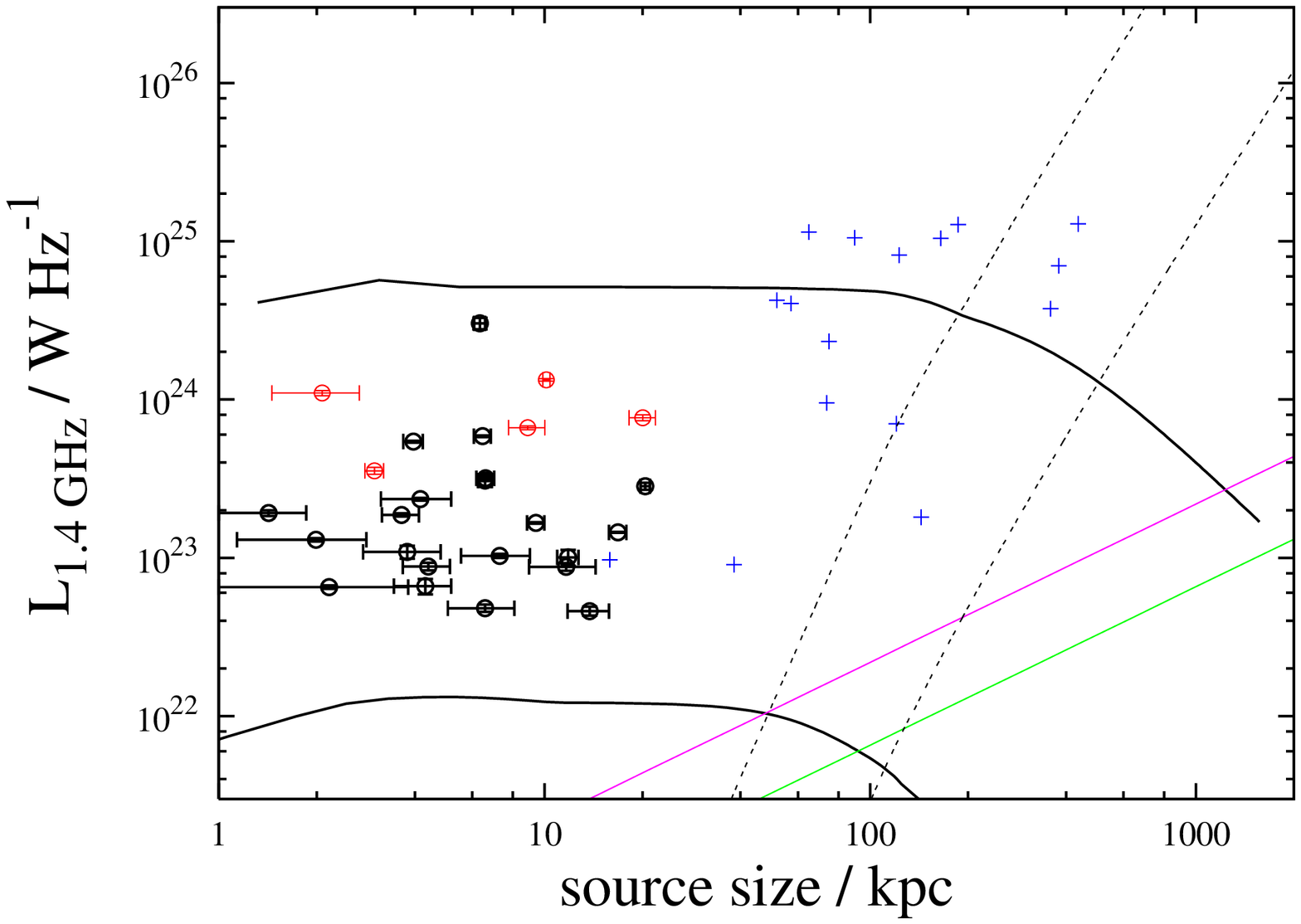}}
	\subfigure[$11.56 < {\rm log} \Mstar\/ / M_{\odot} \leq 11.76$]{\includegraphics[height=0.48\textwidth,angle=270]{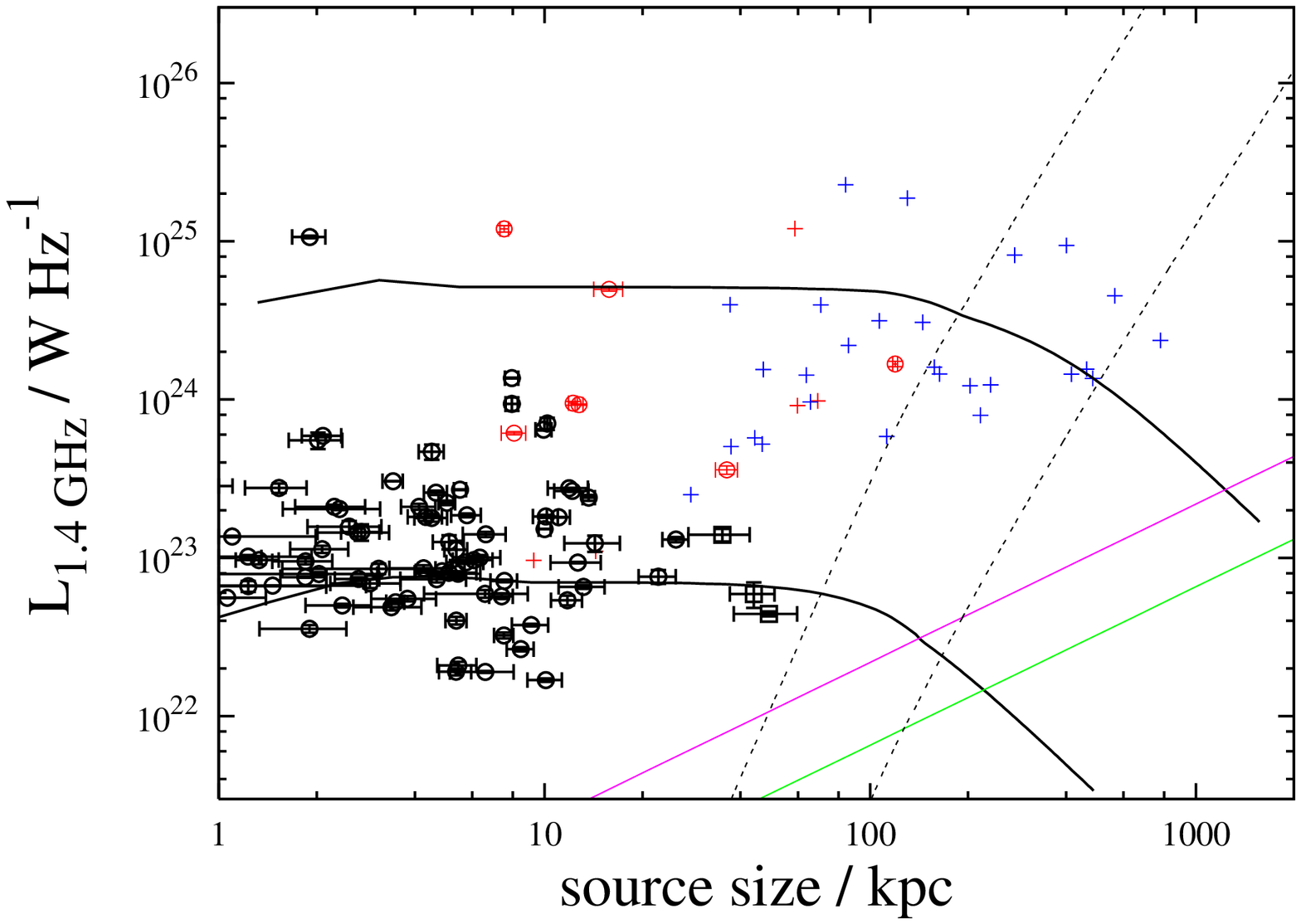}}
	\subfigure[$11.36 < {\rm log} \Mstar\/ / M_{\odot} \leq 11.56$]{\includegraphics[height=0.48\textwidth,angle=270]{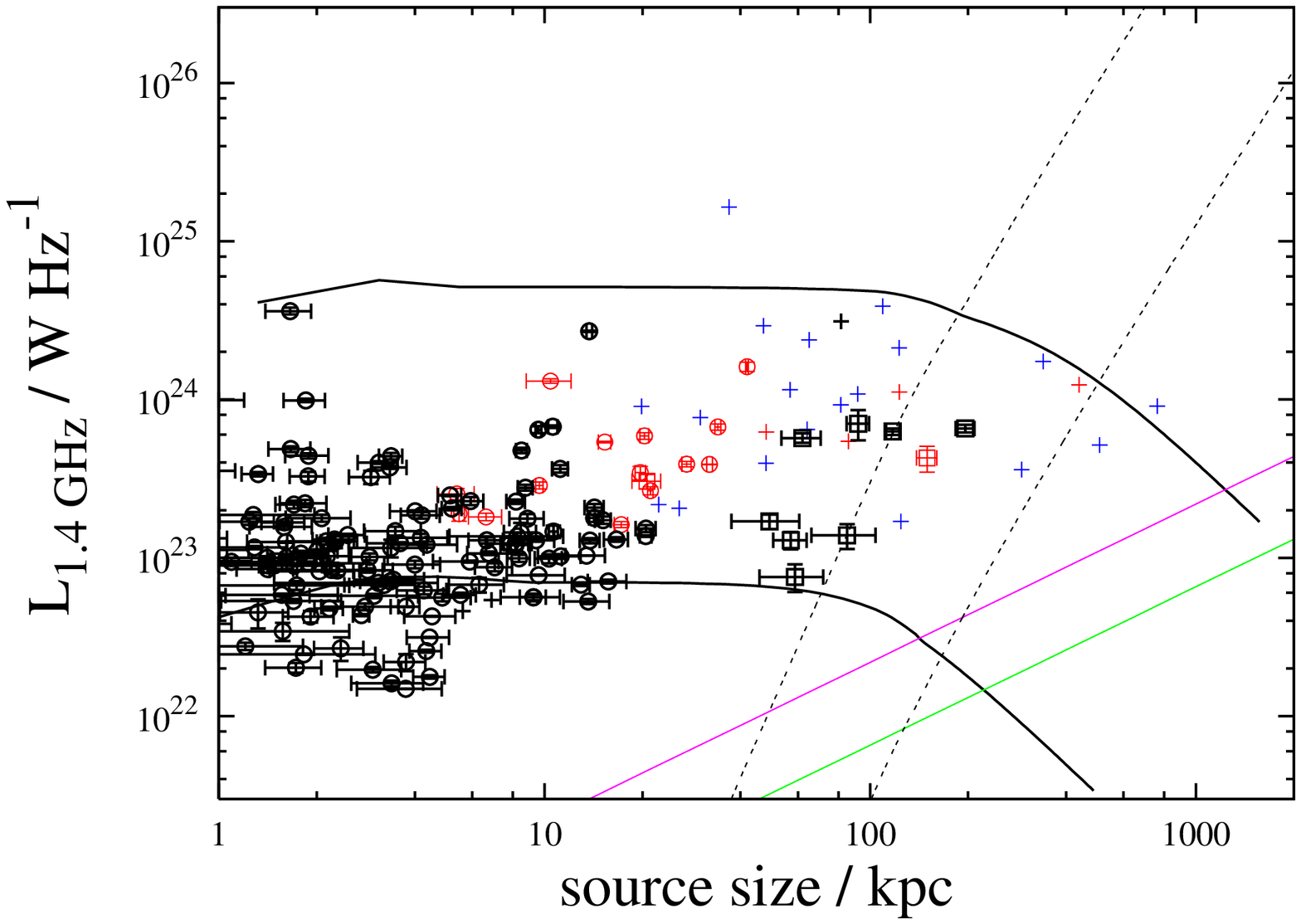}}
	\subfigure[$11.16 < {\rm log} \Mstar\/ / M_{\odot} \leq 11.36$]{\includegraphics[height=0.48\textwidth,angle=270]{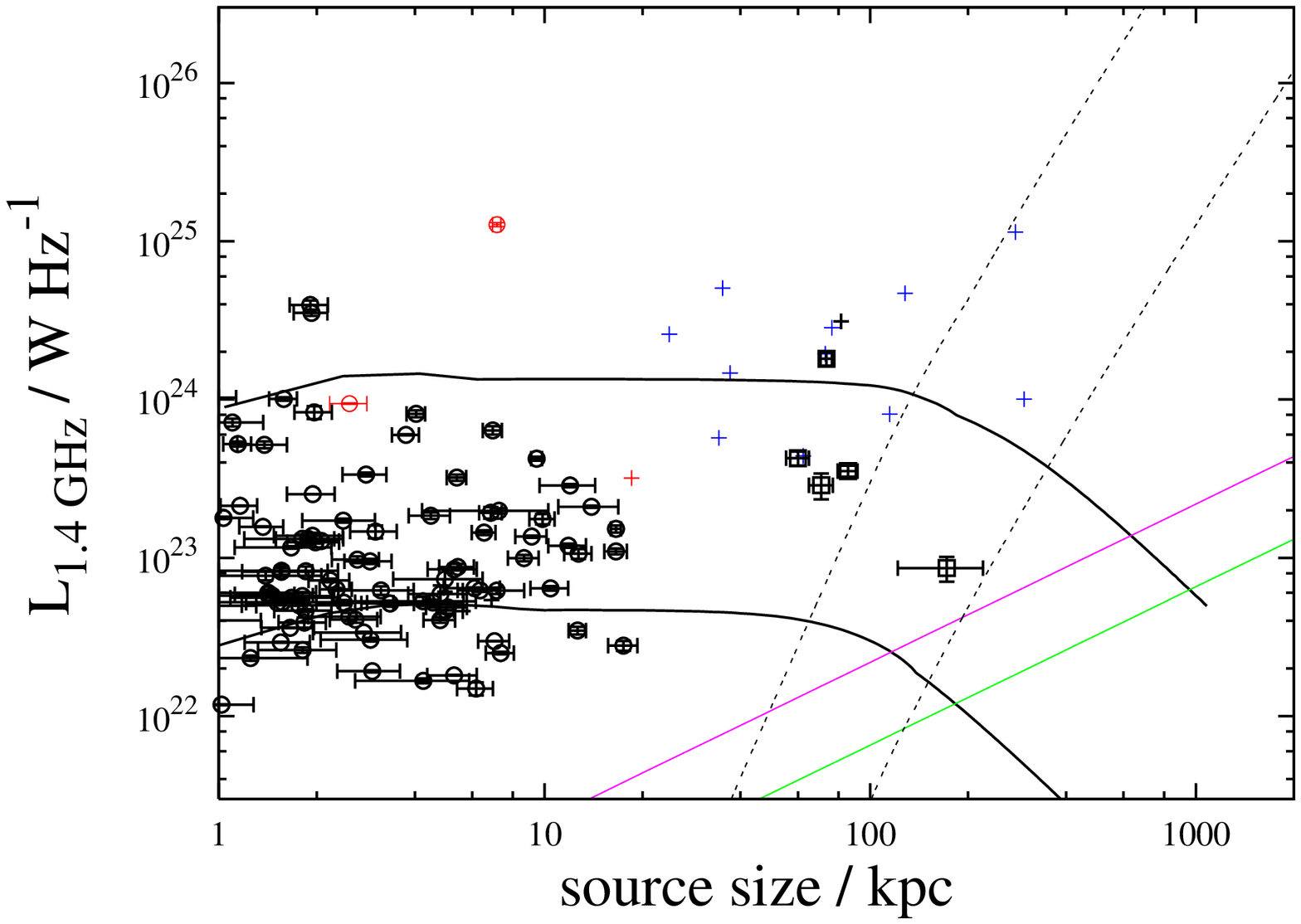}}
	\subfigure[$10.96 < {\rm log} \Mstar\/ / M_{\odot} \leq 11.16$]{\includegraphics[height=0.48\textwidth,angle=270]{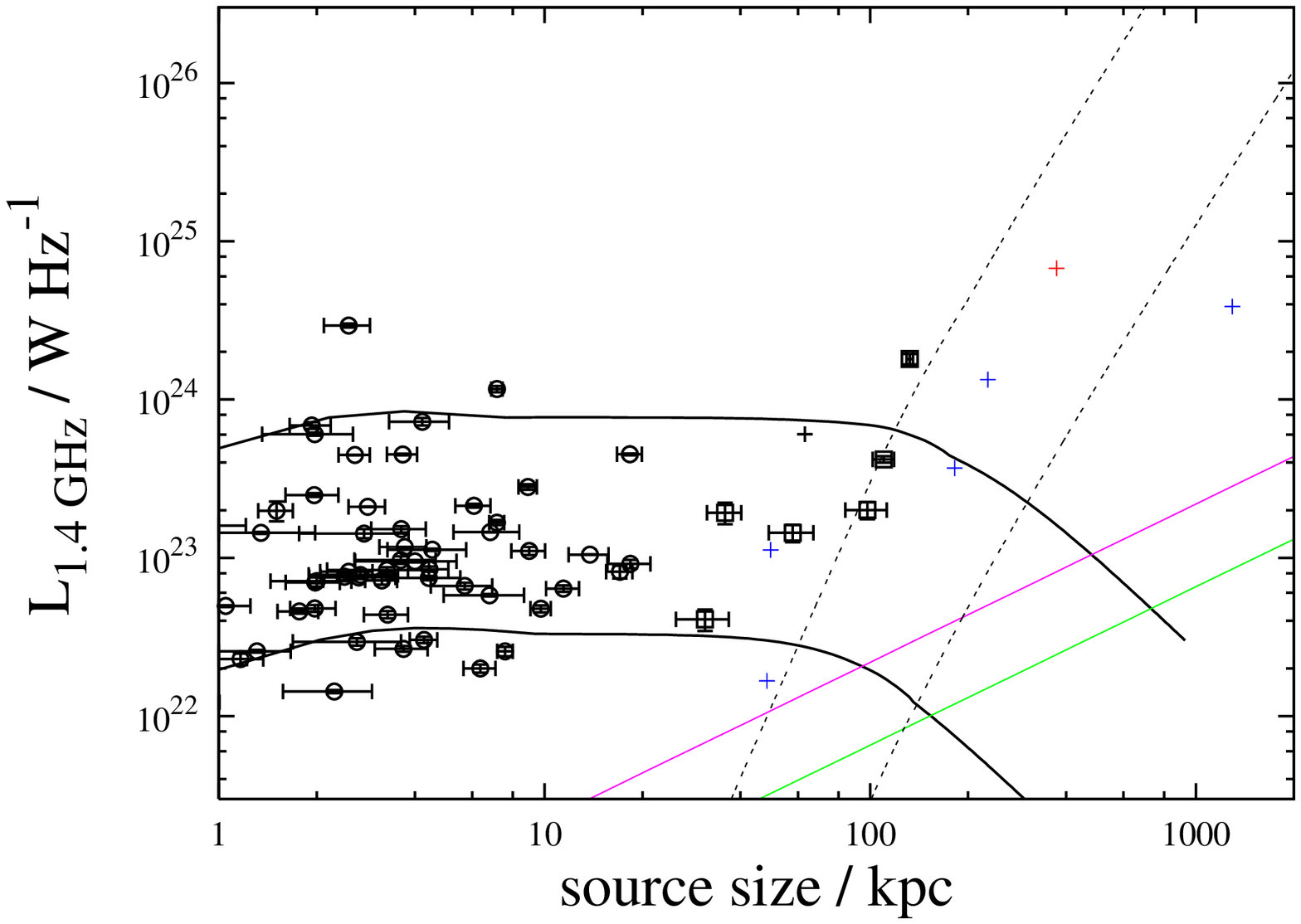}}
\caption{Distribution of radio loud sources in the physical size - radio luminosity plane for a given range of stellar masses. Crosses represent sources whose size is determined by inspection. Open squares represent NVSS, and circles are FIRST sources. Blue points are FR-IIs, red FR-Is; and black points are sources that are only just resolved, and hence have uncertain morphologies. Errors on resolved source sizes are FWHM. Also plotted are the limits on maximum detectable source size, given the sensitivity of each survey, as a function of redshift and luminosity. These are shown as green (NVSS) and purple (FIRST) lines. P-D tracks using parameters of Section~\ref{sec:parameterSpace} with jet power corresponding to mean~$\pm$~$\sigma$ (Table~\ref{tab:derivedParameters}) are also shown as thick black lines. Finally, we plot the expected tracks corresponding to source ages of $10^8$ and $3 \times 10^8$~years; these are shown by dotted lines, with smaller source sizes corresponding to younger sources.}
\label{fig:observedSizesPD}
\end{figure*}

The distribution of sources in the P-D plane can also be used to place constraints on radio source ages. Figure~\ref{fig:observedSizesPD} shows the observed distribution for each $\Mstar\/$ bin. Also plotted are P-D tracks for $\log \Qjet\/ = \log \bar{Q}_{\rm jet} \pm \sigma_{\rm log Q}$ using the parameters of Section~\ref{sec:parameterSpace} and the derived distribution of jet powers given in Table~\ref{tab:derivedParameters}. Curves corresponding to source ages of $10^8$ and $3 \times 10^8$~years are given for guidance; these are the dotted, almost vertical, lines. Finally, limits on the maximum detectable size of a source with given luminosity and redshift are plotted as straight coloured lines in log-log space. Very dim extended sources, such as radio relics, would fall below these limits. As discussed in Section~\ref{sec:selectionEffects}, these are largely missed by the pairing process employed in constructing the present catalogue. However, since in this work the emphasis is on the currently active radio sources, this selection effect does not alter the analysis. More importantly, Figure~\ref{fig:observedSizesPD} shows that there are no resolved sources (i.e. those sources whose positions are well-defined in the P-D plane) close to the NVSS detection threshold (green line). Thus it is unlikely that a significant number of large radio sources is missed.

\begin{figure}
  \centering
  \psfrag{xLabel}{fractile}
  \psfrag{yLabel}{source age / yr}
  \includegraphics[height=0.48\textwidth,angle=270]{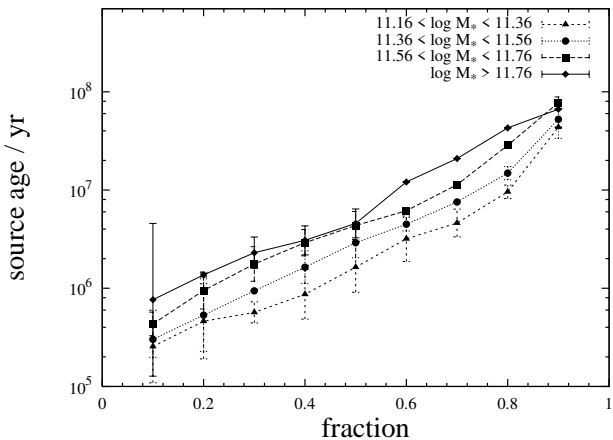}
  \caption{Fractile distribution of radio source ages for the top four stellar mass bins. Lowest mass bins are not shown due to large numbers of unresolved sources in these affecting the statistics significantly.}
  \label{fig:fractiletOn}
\end{figure}

Inspection of Figure~\ref{fig:observedSizesPD} shows that although large (and hence old) sources are found in all mass bins, the median source size is smaller in lower mass bins. Figure~\ref{fig:fractiletOn} quantifies this statement by using the derived source ages and once again plotting the fractile ranges. Unresolved sources pose a problem for this approach, since only upper limits on their sizes are available. We assume that all unresolved sources have very small sizes and are thus younger than the smallest resolved sources. While crude, this assumption is sufficiently accurate for the highest stellar mass bins, where the unresolved number counts are low, and there are many sources larger than the upper unresolved limit. However, no meaningful information can be extracted for ${\rm log}~\Mstar\/ / M_{\odot} \leq 11.16$.

Much like jet powers, the derived source ages appear to depend strongly on stellar mass, with the most massive galaxies hosting older sources. Using the median $\tOn\/$ values (Table~\ref{tab:derivedParameters}) we find $\tOn\/ \propto \Mstar\/^{1.2}$.

\subsection{Quiescent timescales}
\label{sec:tOff}

\begin{figure*}
\centering
	\subfigure[${\rm log} \Mstar\/ / M_{\odot} > 11.76$]{\includegraphics[height=0.4\textwidth,angle=270]{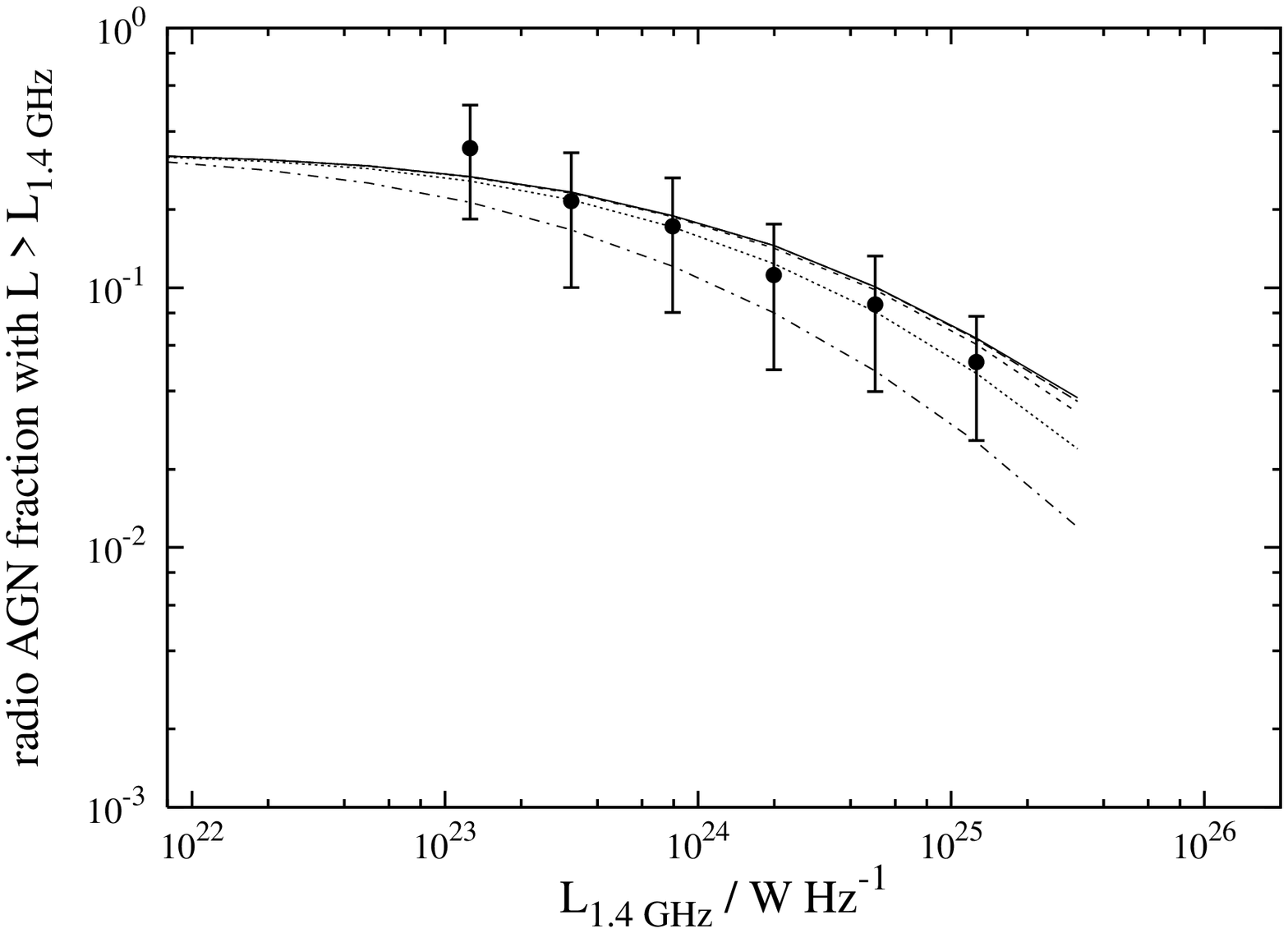}}
	\subfigure[$11.56 < {\rm log} \Mstar\/ / M_{\odot} \leq 11.76$]{\includegraphics[height=0.4\textwidth,angle=270]{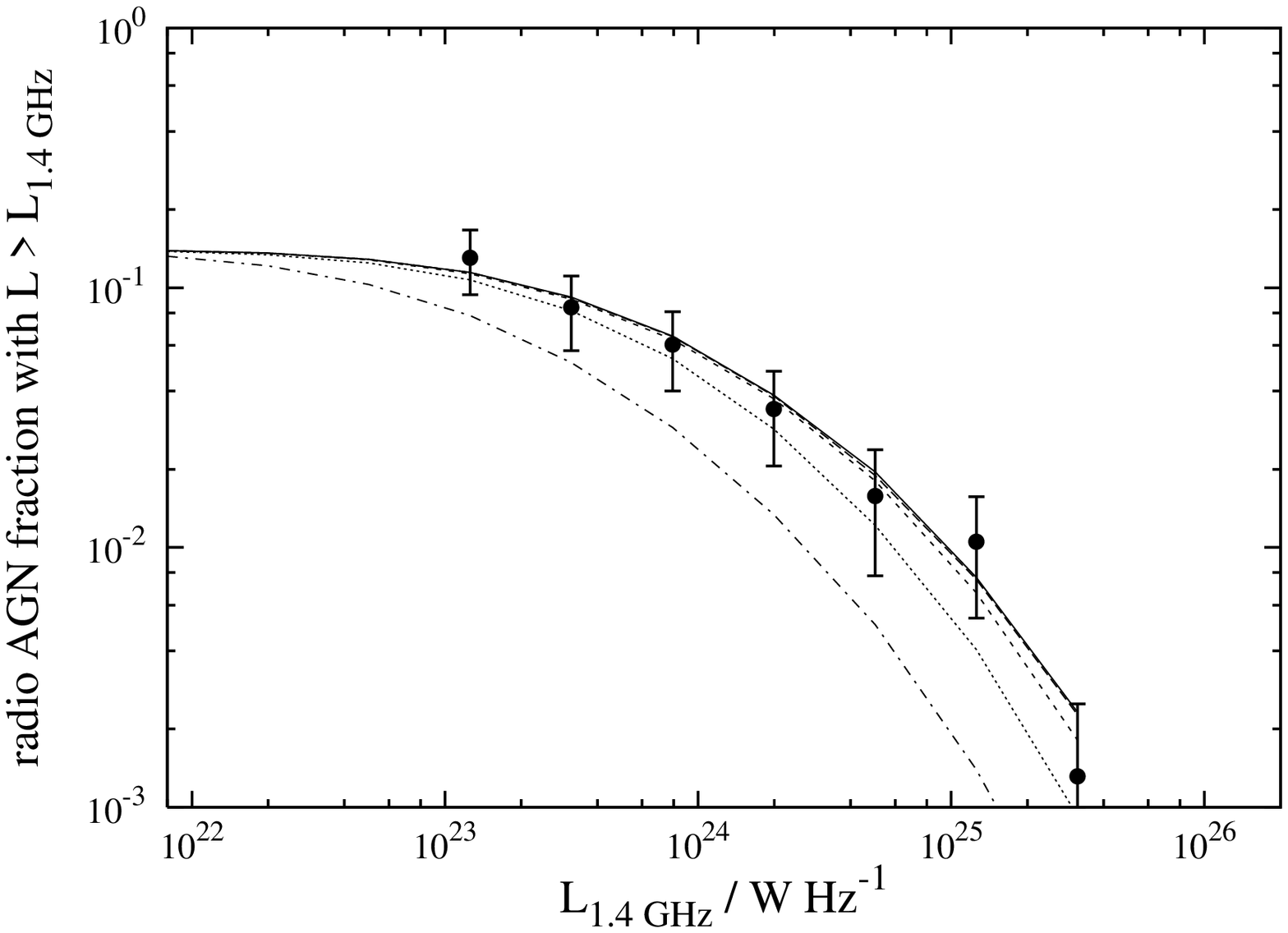}}
	\subfigure[$11.36 < {\rm log} \Mstar\/ / M_{\odot} \leq 11.56$]{\includegraphics[height=0.4\textwidth,angle=270]{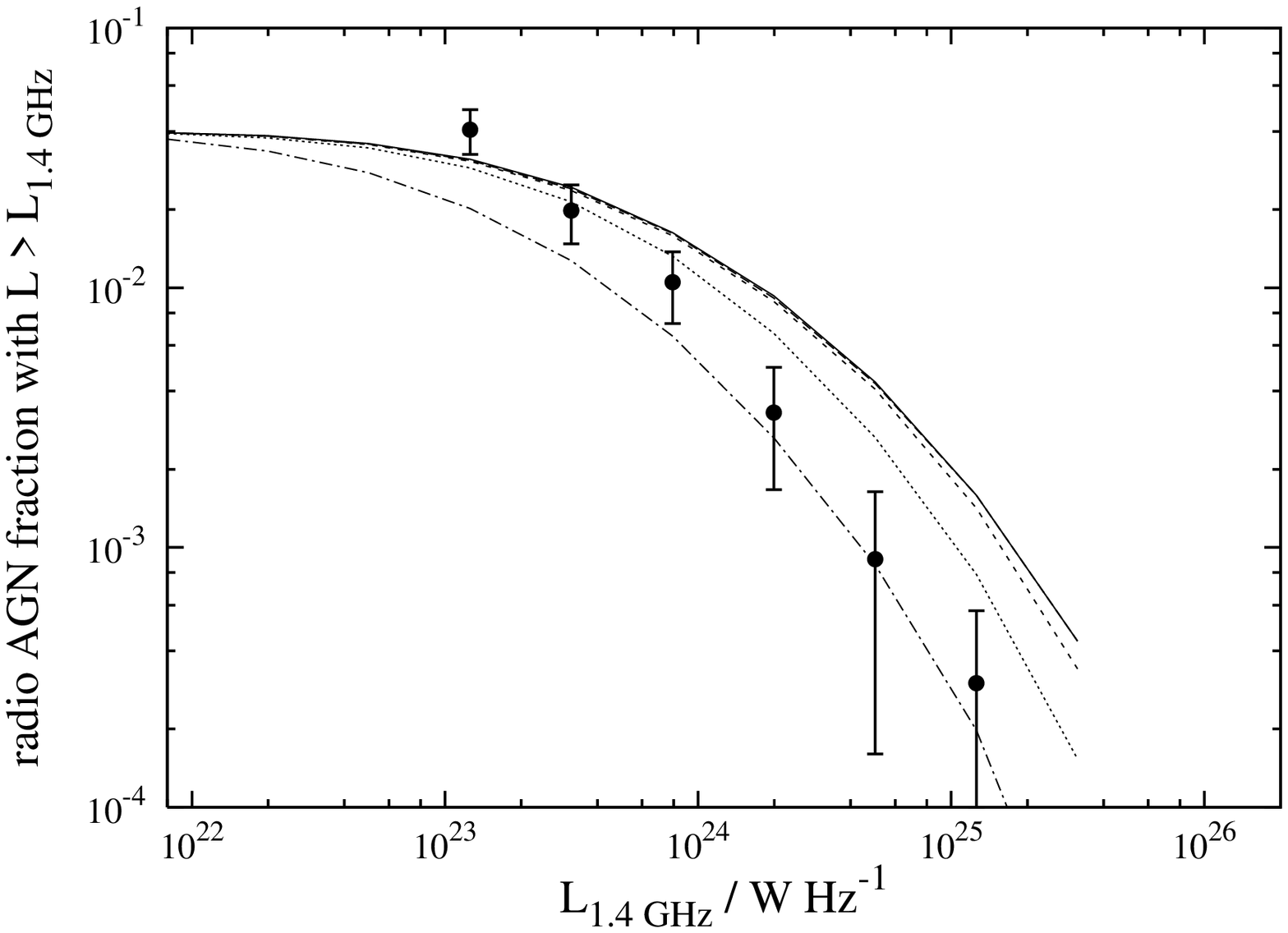}}
	\subfigure[$11.16 < {\rm log} \Mstar\/ / M_{\odot} \leq 11.36$]{\includegraphics[height=0.4\textwidth,angle=270]{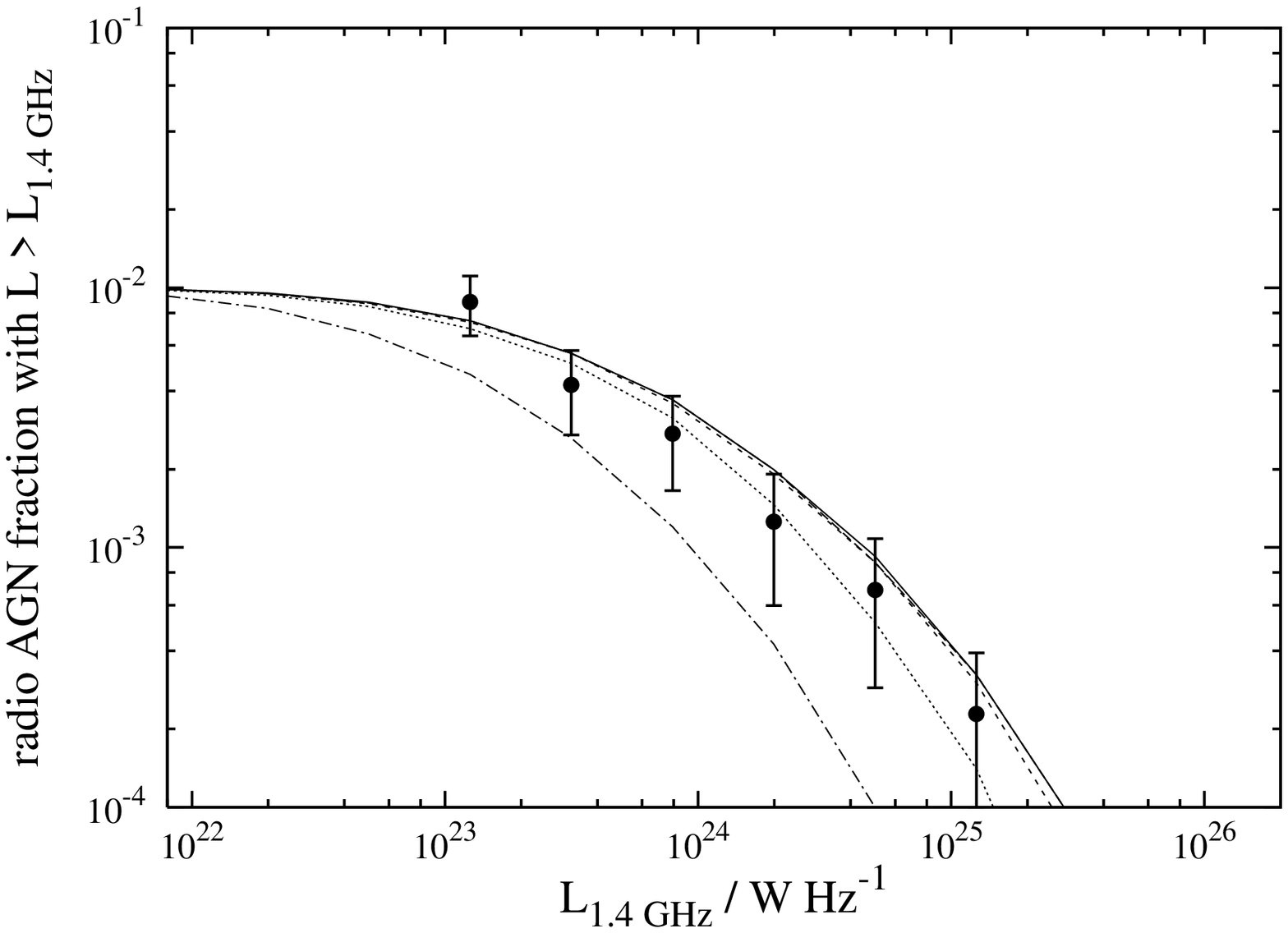}}
\caption{Predicted cumulative radio luminosity functions as a function of stellar mass. The curves represent various jet on times; solid = $10^7$~yrs, long-dashed = $2.5 \times 10^7$~yrs, short-dashed = $8 \times 10^7$~yrs, dotted = $2 \times 10^8$~years, dot-dashed = $6.3 \times 10^8$~years. Parameters used for fitting are the same as those from which source ages and jet powers were determined. The observed luminosity functions are only plotted for $L_{\rm 1.4}>8 \times 10^{22}$~\WHz\/ as the sample is incomplete below this luminosity. No plot is shown for ${\rm log} \Mstar\/ / M_{\odot} < 11.16$ due to large uncertainties in derived jet powers.}
\label{fig:fittedBLFs}
\end{figure*}

Duration of the quiescent phase $\tOff\/$ is found from the bivariate luminosity function. Figure~\ref{fig:fittedBLFs} plots the best fits to the observed RLF for each stellar mass bin. Parameters of Sections~\ref{sec:parameterSpace} and~\ref{sec:sourceSizes}, and jet powers given in Table~\ref{tab:derivedParameters}, are adopted. The only adjustable parameters are the jet on and off times.

As discussed in Section~\ref{sec:parameterSpace}, it is difficult to constrain the active phase duration from the radio luminosity function alone. However, the RLFs are very sensitive to the ratio $\tOff\/ / \tOn\/$. Therefore, combining $\tOn\/$ estimates from observed source sizes with constraints on the duty cycle from the luminosity functions yields $\tOff\/$ values for each $\Mstar\/$ bin. These are given in Table~\ref{tab:derivedParameters}, where $\tOn\/$ is calculated from the median values of Table~\ref{tab:derivedParameters} by assuming that a typical radio source with a given stellar mass will be active for $2 t_{\rm on, median}$, i.e. it is observed halfway through its evolution.

\section{Discussion}
\label{sec:discussion}

\subsection{Robustness of timescale estimates}
\label{sec:timescaleRobustness}

It is important to consider how reliable the timescales derived in the preceding section are. The ratio $\tOff\//\tOn\/$ ratio is largely set by the radio loud fraction in each bin (see Section~\ref{sec:parameterSpace}), and hence uncertainties in this quantity are comparable with Poisson errors in the corresponding luminosity function. These are clearly negligible in comparison with variability in parameters such as the density profile, jet power and cocoon axial ratio, all of which are important for individual source age determination.

The dynamical model of Kaiser \& Alexander \shortcite{KA97} relates the size of a source expanding in a power-law atmosphere with exponent $\beta$ to its age $t$ and various cocoon and ICM parameters via

\begin{equation}
  D \propto \RT\/^{-\frac{4}{5-\beta}} \left( \frac{\Qjet\/ t^3}{\rhoCore\/ \rCore\/^{\beta}} \right) .
\label{eqn:sourceSize}
\end{equation}

Here, the cocoon axial ratio $\RT\/$ is inversely proportional to jet opening angle $\theta$. For the majority of our sample we do not expect $\theta$ to vary by more than a factor of two, and as $t \propto \RT\/^{-4/3}$ we do not expect the axial ratio to affect our source age determinations by more than about a factor of two also.

Neglecting relativistic electron energy loss processes (adiabatic and radiative), Kaiser \etal\/ \shortcite{KDA97} model gives 
\begin{equation}
  \Lradio\/ \propto (\rhoCore\/ \rCore\/^\beta)^{\frac{5+\beta}{12}} \Qjet\/^{\frac{5+p}{6}} D^{\frac{16-4p-(p+5)\beta}{12}} . \nonumber \\
\end{equation}

where $p=2.14$ is the power-law exponent of the electron energy distribution. Rearranging yields

\begin{equation}
  \Lradio\/ \propto \Qjet\/^x (\rhoCore\/ \rCore\/^\beta)^y t^{\frac{3}{5-\beta}}
\label{eqn:Lradio}
\end{equation}

where $x=\frac{66+6p-3(p+5)\beta}{12(5-\beta)}$ and $y=\frac{(9+4p)+\beta (p+5-\beta)}{12(5-\beta)}$. In deriving individual source jet powers in Section~\ref{sec:sourceSizes} we had to assume a density profile. However, from Equation~\ref{eqn:sourceSize} it is clear that source size does not depend on $\Qjet\/$ or the density profile individually, but instead the combination $q=\frac{\Qjet\/}{\rhoCore\/ \rCore\/^{\beta}}$. By knowing both the source radio luminosity and size, and using equations~\ref{eqn:sourceSize} and \ref{eqn:Lradio} we arrive at $t \propto q^{-\left( \frac{x+y}{3y+\left( \frac{3}{5-y} \right)} \right)}$. Taking $\beta=1.5$ gives $x=1.1$ and $y=0.6$, yielding $t \propto q^{-0.5}$. Hence even an uncertainty of an order of magnitude in $q$ will alter the derived timescale by only a factor of a few. We can thus conclude that our $\tOn\/$ estimates are likely to be correct to within a factor of two or so. As the uncertainty associated with the ratio $\tOff\//\tOn\/$ is much less than this, duration of the quiescent timescale is also expected to be similarly accurate.

\subsection{Mass dependence}
\label{sec:massDependence}

Best \etal\/ \shortcite{Paper2} found a strong dependence of the radio loud fraction on black hole and host galaxy mass, $f_{RL} \propto \Mbh\/^{1.6 \pm 0.1}$. We find a similar result for our volume-limited sample (Figure~\ref{fig:BLF}), with $f_{RL} \propto \Mstar\/^{2.1 \pm 0.3}$. For a fixed spheroid-to-total light ratio (Section~\ref{sec:jetPowers}), using the $\Mbh\/-\Mstar\/$ relation \cite{HaeringRix04} gives $f_{RL} \propto \Mbh\/^{1.8 \pm 0.5}$. Best \etal\/ \shortcite{Paper2} point out that this relation is similar to the $\MdotCool\/ \propto \Mbh\/^{1.5}$ gas cooling rate dependence on black hole mass derived from the X-ray-optical correlation in luminous elliptical galaxies, and suggest that AGN radio activity might be fuelled by the cooling of hot gas within the host galaxy.

The derived AGN ``on'' timescales in Table~\ref{tab:derivedParameters} clearly increase with stellar mass. If the AGN is fuelled by cool gas accretion, and the duration of the active phase is limited by fuel availability, one would expect for constant radiative and accretion efficiency $\tOn\/ \propto \frac{\MdotCool\/}{\Qjet\/}$, which on using the scaling relations of Section~\ref{sec:jetPowers} gives $\tOn\/ \propto \Mstar\/^{0.9 \pm 0.5}$. This is consistent with $\tOn\/ \propto \Mstar\/^{1.2}$ derived in Section~\ref{sec:tOn}, and suggests that the duration of the active phase may indeed be determined by the availability of fuel to ``feed'' the jets. The quiescent phase duration (Table~\ref{tab:derivedParameters}) also shows a strong dependence on $\Mstar\/$, with sources hosted by the most massive galaxies spending a shorter amount of time in the off phase, consistent with higher cooling rates in those objects. In the absence of a heating source one would expect $\tOff\/ \propto \MdotCool\/^{-1} \propto \Mstar\/^{-1.7 \pm 0.3}$. However, in practice as the more massive galaxies host the more powerful radio sources, they will be more affected by the heating and cold gas uplifting than their lower mass counterparts, and the observed relation is not as steep.

\section{Conclusions}
\label{sec:conclusions}

We constructed a flux- and volume-limited ($0.03 \leq z \leq 0.1$) sample of radio sources with optical identifications by cross correlating the SDSS optical survey with 1.4 GHz NVSS and FIRST surveys. Source sizes and luminosities together with radio source models allowed us to derive jet powers and ages of individual sources, and hence the jet ``on'' time as a function of stellar mass. The bivariate luminosity function was then used to constrain the time a typical radio source spends in an inactive state. Radio and emission line AGN activity are found to be independent phenomena. We also find that both the radio source lifetime and duration of the quiescent phase have a strong mass dependence, with massive hosts harbouring longer-lived sources that are triggered more frequently. Gas cooling rate shows a similar mass dependence, suggesting that fuel depletion is the reason the jets switch off.

\section*{Acknowledgements}

We thank the Commonwealth Cambridge Trust (SS) and the Isaac Newton Trust (SS and SA) for support, and the anonymous referee for comments that have helped improve the paper. This work makes use of the SDSS Archive, funding for which has been provided by the Alfred P. Sloan Foundation, the Participating Institutions, the National Science Foundation, the U.S. Department of Energy, the National Aeronautics and Space Administration, the Japanese Monbukagakusho, the Max Planck Society, and the Higher Education Funding Council for England. This work also makes use of the NVSS and FIRST surveys carried out using the National Radio Astronomy Observatory Very Large Array. The National Radio Astronomy Observatory is a facility of the National Science Foundation operated under cooperative agreement by Associated Universities, Inc.

\end{document}